\documentclass[preprint,aps,superscriptaddress,showkeys,11pt]{revtex4}
\pdfoutput=1
\usepackage{graphicx}
\usepackage{amsmath}
\usepackage{amsfonts}
\usepackage{amssymb}
\usepackage{xcolor, soul}
\usepackage{epstopdf}
\usepackage{float}
\usepackage{subfigure}
\usepackage{hyperref}
\hypersetup{
	colorlinks   = true,
	citecolor    = red,
	linkcolor    = blue,
	urlcolor     = blue,
}
\newcommand{\be}{\begin{equation}}
\newcommand{\ee}{\end{equation}}
\newcommand{\bea}{\begin{eqnarray}}
\newcommand{\eea}{\end{eqnarray}}

\allowdisplaybreaks

\graphicspath{{./figs/}}

\begin{document}
	\title{Effective Field Theory of Hairy Black Holes and Their flat/dS limit}
	\author{Sayan Chakrabarti}%
	\email{sayan.chakrabarti@iitg.ac.in}
	\affiliation{%
		Department of Physics, Indian Institute of Technology Guwahati.\\
		Guwahati, Assam, India }%
	\author{Sumeet Chougule}%
	\email{chougule@cecs.cl}
	\affiliation{%
		Department of Physics, Indian Institute of Technology Guwahati.\\
		Guwahati, Assam, India }%
	\affiliation{%
	Centro de Estudios Cient\'{\i}ficos (CECs), Av. Arturo Prat 514, Valdivia, Chile}%
	\affiliation{%
		Departamento de F\'{\i}sica, Universidad de Concepci$\acute{o}$n,\\
		 Casilla 160-C, Concepci$\acute{o}$n, Chile}%
	\author{Debaprasad Maity}%
	\email{ debu@iitg.ac.in}
	\affiliation{%
		Department of Physics, Indian Institute of Technology Guwahati.\\
		Guwahati, Assam, India }%


	%
	\begin{abstract}
		Effective theory of fluctuations based on underlying symmetry plays a very important role in understanding the low energy phenomena. Using this powerful technique we study the fluctuation dynamics keeping in mind the following central question: does the effective theory of black hole provide any information about the possible existence of hair? Assuming the symmetry of the hair being that of the underlying black hole space-time, we start by writing down the most general action for the background and the fluctuation in the effective field theory framework. Considering the Schwarzschild and Schwarzschild de Sitter black hole background with a spherically symmetric hair we derived the most general equation of motion for the fluctuation. For a particular choice of theory parameters, quasinormal modes corresponding to those fluctuations appeared to have distinct features compared to that of the usual black hole quasinormal modes. The background equations from the effective theory Lagrangian, on the other hand, seemed to suggest that the underlying theory of the hair under consideration should be higher derivative in nature. Therefore as a concrete example, we construct a class of higher derivative scalar field theory which gives rise to spherically symmetric hair through background cosmological constant. We also calculate the quasinormal modes whose behaviour turned out to be similar to the one discussed from the effective theory.       
	\end{abstract}
	
	\maketitle
	
	\tableofcontents
	
	\section{Introduction}
	General theory of relativity has been proved to be very successful in explaining a large variety of gravitational phenomena in a wide range of length scales. The theory also provides us with a class of vacuum solutions which are called black holes. Black holes are the most fascinating objects in general relativity which has not been clearly understood from different aspects despite having a long history of various path-breaking researches. On the other hand, it is believed to be the simplest object in our universe, which is characterised by only three parameters: mass, charge and angular momentum. Over the years, it has also been understood that those are the only three charges which a black hole can carry. In black hole physics, this is known as black hole no-hair theorem \cite{nohair,nohair2}. A large amount of work has been done in the literature on the possible existence of hairy black holes for various different kind of theories. Let us particularly mention a class of modified gravity theories \cite{modified} which has gained significant attention in the recent past because of their applicability in various physical contexts. The existence of black hole hairs has been discussed in those type of theories quite extensively in \cite{hui,TS2013,other}. We will also discuss a particular class of these theories in the present paper. Having talked about an extensive research area in the domain of black hole physics, it has to be admitted that all those endeavours towards understanding black holes are mostly confined within the theoretical regime, without much to do with the observation. Thanks to the recent breakthrough on the observational front in gravitational waves \cite{TheLIGOScientific:2017qsa,Abbott:2017gyy} originating from the merger of supermassive black holes in a binary, has finally opened up the exciting possibility of verifying as well as understanding those large volume of theoretical works and most importantly in understanding the deeper underlying principles of the nature of space time. All the studies on black holes so far were model dependent, where one considers a specific theory of gravity and its black hole solutions. On the other hand, when it comes to observations, all the observable quantities are based upon the properties of fluctuations around the black hole background without much to do with a specific theory of gravity. This motivates us to understand black holes in an effective field theory framework. As the name suggests, it will be a model-independent description of fluctuations around a black hole space time. In this paper, our analysis will be focused on the following question: {\em does the effective theory of black holes provide any information about the possible existence of hair?} Our study of possible hairy black hole solutions will be valid for a class of asymptotically flat and dS black holes. As has been pointed out before, black holes with hair is a subject of intense research for the past many years. For a very interesting short review, the reader is referred to \cite{radu} and {\em references therein}. 
	
	The effective field theory approach has been proved to be a universal tool to understand low energy phenomena in many areas of physics. One particularly interesting application of this approach, which has gained significant attention in the recent past, is the effective theory of inflation \cite{EFT}. The idea behind this theory was to understand the model independent dynamics of fluctuation in a time-dependent inflaton background. The most important ingredient is the spontaneous breaking of time translation symmetry by the inflaton. Motivated by this, we initiated the present work. Our basic underlying assumption in this work would be to consider the existence of a hairy black hole which breaks spatial translation symmetry. For a generic theory, a non-trivial background solution always breaks a certain amount of symmetry of the underlying theory. For a theory of gravity, diffeomorphism is the basic underlying symmetry. The existence of non-trivial black hole hair will break the aforementioned symmetry. Therefore, the main idea in our approach will be to write down all possible terms in the effective Lagrangian which obey the residual diffeomorphism symmetry in the above mentioned hairy black hole background. 

For our present study, we will consider the simplest situation where the background is a spherically symmetric black hole with the hair having the same symmetry. The origin of the hair in the effective field theory framework is not important for our study. More general situations will be discussed in our future publications. Before we embark on our analysis, let us mention some independent studies on fluctuations in the black hole background. In reference \cite{tsujikawa}, the authors have studied the fluctuation dynamics in static black hole background considering the less symmetric situation with angular diffeomorphism invariance. This, in principle, will lead to more than one degrees of freedom as opposed to the case considered in our current paper for a hairy black hole. In another work \cite{lagos}, more elaborate formalism has been developed irrespective of specific hairy background and it is valid even for non-hairy black holes. However, to achieve the full diffeomorphism invariant action the formalism requires to start with a large number of independent parameter and then systematic analysis needs to be performed to finally reduce the number of parameters. While we were studying properties of fluctuations in a hairy black hole background based on the formalism used in effective theory of inflation, the paper \cite{Franciolini:2018uyq} came up where the same formalism has been considered and it studied the properties in greater detail.
	
We organise our paper as follows: in section-II, we write down the most general Lagrangian for the background and the associated fluctuations assuming  $r$-diffeomorphism symmetry to be broken by some unknown field. Our goal is to understand the hairy black holes in the effective field theory framework. Therefore we will assume a spherically symmetric black hole background with a hair which also inherits the symmetry of the background. By using the well known St\"{u}ckelburg mechanism, we identify the scalar Goldstone mode associated with the broken $r$-diffeomorphism symmetry and derive its equation of motion. While deriving the equation, the background naturally plays a very important role. Therefore, fluctuation equation encodes the information about the nature of hair in the background. Considering the simplest asymptotically flat/dS/AdS black hole backgrounds, our effective field theory approach shows that they can support hair which generates background cosmological constant dynamically. However, it turns out that the theory of the hair, in general, should be higher derivative in nature. This fact may be intimately tied with the conventional no-hair theorem, which will be discussed in the future. Considering this observation, we will construct a specific higher derivative theory of scalar field in the subsequent section. This section is ended by discussing the qualitative behaviour of the quasinormal modes for the Goldstone modes which behaves distinctly as opposed to the conventional black hole quasinormal modes. Interestingly enough, this distinct behaviour also can be seen from our underlying theory discussed in detail in section-III. In this section, we will discuss a class of higher derivative theory of the scalar field. The vacuum expectation value of the scalar breaks $r$-diffeomorphism symmetry with a background cosmological constant. With this background, we study the behaviour of quasinormal modes. Finally, we finish the paper with the conclusion and future directions.

	\section{Effective field theory setup}
 
	In this section, we will formulate the theory. As already mentioned, we assume there exists a hairy black hole with spherical symmetry. It naturally breaks the $r$-diffeomorphism symmetry. Therefore, without requiring any specific model of gravity, we can write down the most generic Lagrangian which will be invariant under the residual diffeomorphism symmetry. Under the residual symmetry transformation ${x^{\prime}}^i = x^i + \xi^i$, the list of covariant quantities contains $g^{rr}$ and all the geometrical objects defined on an arbitrary $r={\rm constant}$ hypersurface, namely the extrinsic curvature $K_{\mu \nu}$, three-dimensional Ricci scalar $\mathcal{R}^{(3)}$. However, one important point we missed is the degree of symmetry of the $r$-constant hypersurface which is not maximally symmetric. The spatially flat cosmological background is one example where $t$-constant hypersurface is maximally symmetric. Therefore, for any spherically symmetric background, the extrinsic curvature $K_{\mu\nu}$ of $r$-constant hyper-surface does not become proportional to the induced metric. As a result, one needs to consider the rotational symmetry and general time reparametrisation invariance separately as pointed out and discussed in detail in the reference \cite{Franciolini:2018uyq}. Therefore, in the framework of effective field theory, we need a large number of independent parameters compared to that of the extremal case to understand the dynamics of fluctuation in the hairy black hole background. With these important points and ingredients, the most generic Lagrangian for the background and the fluctuation can be expressed as
	\begin{align} \label{a}
	\mathcal{S}= \int d^{4}x \sqrt{-g}\Bigg\{&\frac{M_{Pl}^{2}}{2}\mathcal{R}-\Lambda(r) - c(r)g^{rr} -\alpha(r) \bar{K}_{\mu\nu} K^{\mu\nu}+\frac{1}{2!}M_{2}(r)^{4} (\delta g^{rr})^{2}  +\frac{1}{3!}M_{3}(r)^{4}(\delta g^{rr})^{3} \nonumber \\
	&-\frac{\bar{M}_{1}(r)^{3}}{2}(\delta g^{rr})(\delta K^{\mu}\,_{\mu})    + \bar{M}_{4}^{2}(r) \bar{K}_{\mu\nu}\delta g^{rr} \delta K^{\mu\nu} +M_{5}^{2}(r) (\partial_{r}\delta g^{rr})^{2} + M_{6}^{2}(r) (\partial_{r}\delta g^{rr}) \delta K^{\beta}\,_{\beta} \nonumber \\ 
	& + M_{7}^{2}(r) \bar{K}_{\alpha \beta} (\partial_{r}\delta g^{rr}) \delta K^{\alpha \beta} + \frac{\hat{m}^{2}_{2}(r)}{8} g^{\mu \nu} \partial_{\mu}(\delta g^{rr}) \partial_{\nu} (\delta g^{rr}) -\frac{\bar{M}_{2}(r)^{2}}{2}(\delta K^{\mu}\,_{\mu})^{2}  \nonumber \\
	&   -\frac{\bar{M}_{3}(r)^{2}}{2}\delta K^{\mu}\,_{\nu}\delta K^{\nu}\,_{\mu}+ M_{11}(r) \bar{K}_{\mu\nu}\delta K^{\beta}\,_{\beta} \delta K^{\mu\nu} + M_{12}(r) \bar{K}_{\mu\nu}\delta K^{\mu\rho} \delta K^{\nu}\,_{\rho} \nonumber \\
	&  + \lambda_{1}(r) \bar{K}_{\mu\rho}\bar{K}^{\rho}\,_{\nu}\delta K^{\beta}\,_{\beta} \delta K^{\mu\nu}  -\frac{\hat{m}^{2}_{1}(r)}{2}\delta g^{rr}\delta \mathcal{R}^{(3)} + M_{14}(r) \bar{K}_{\mu\nu}\delta g^{rr} \delta R^{\mu\nu(3)}\Bigg\}.
	\end{align}
	All the parameters take care of the dimensions of the corresponding term and all of them are functions of $r$. The background of extrinsic curvature is given by $\bar{ K}_{\mu\nu}=\frac{1}{2}\frac{g'_{\mu\nu} }{\sqrt{g(r)}} $. The above action contains all the possible terms at the quadratic level. The interested reader can find a very good description of why these are the only terms possible in the appendix of  \cite{Franciolini:2018uyq}. In the following discussion, we will keep up to the fourth derivative order of the fluctuation. For the above general Lagrangian, we consider the following form of the spherically symmetric hairy black hole background:
		\bea \label{eq2}
	ds^2 = -f(r)dt^2 + {g(r)} {dr^2} + R(r)d\Omega^2~~;~~\phi \equiv \phi_0(r).
	\eea
	At this point, let us mention that $\phi$ represents the hair, whose explicit nature is not important to us. Because of $r$-diffeomorphism symmetry of the underlying theory, one can always choose unitary gauge, $\delta \phi = 0$ such that the extra scalar degree of freedom is eaten in the metric and manifest itself as a gravitational degree of freedom which we parametrised by the fluctuation of the metric $\delta g^{rr} = g^{rr} - 1/g(r)$, and consequently we can express all the covariant geometric quantities defined on $r$-constant hypersurface in terms of those metric fluctuations. Based on these general arguments, the first four terms in the above action encodes the information about the hairy black hole background given in Eq. (\ref{eq2}). Since the coefficient of arbitrary linear fluctuation is proportional to the equation of motion, the rest of the terms in the Lagrangian will start from quadratic order in fluctuation. Hence unperturbed background will be parametrised by three unknown functions $\Lambda(r)$, $c(r)$ and $\alpha (r)$, and associated stress-energy tensor can be expressed as 
	\begin{align}
	T_{\mu\nu}=&-\frac{2}{\sqrt{-g}}\frac{\delta \mathcal{S}}{\delta g^{\mu\nu}} =
	-g_{\mu\nu} \left[c(r)g^{rr} +\Lambda(r) + \alpha(r) K_{\alpha \beta} K^{\alpha \beta} \right] +2c(r)\delta^{r}_{\mu} \delta^{r}_{\nu}  -  \alpha(r) K_{\alpha \beta} K^{\alpha \beta} n_{\mu} n_{\nu} \nonumber \\ 
	&{- \nabla_{\beta} \big( \alpha(r) K^{\beta}\,_{ \mu} n_{\nu} \big) - \nabla_{\beta} \big( \alpha(r) K^{\beta}\,_{ \nu} n_{\mu} \big) +\nabla_{\beta} \big( \alpha(r) K_{ \mu \nu} n^{\beta} \big) }.
	\end{align}
	By using the Einstein's equation $R_{\mu\nu} - \frac{1}{2} R g_{\mu\nu} =({1}/{M_{Pl}^{2}})T_{\mu\nu}$, the expressions for the two unknown parameters $(\Lambda(r),c(r))$ turn out to be
	\begin{align} \label{eq4}
	\Lambda(r)&=\frac{M_{Pl}^{2}}{g(r)}\bigg(-\frac{f'(r)R'(r)}{4 f(r) R(r)}+\frac{g'(r)R'(r)}{4 g(r) R(r)} + \frac{g(r)}{R(r)} -\frac{R''(r)}{2 R(r)}  \bigg)   \nonumber \\
	&  {-\bigg( \frac{f'^{2}(r)}{2f^{2}(r)} +\frac{f'(r)g'(r)}{4 g^{3/2}(r) \sqrt{f(r)}} -\frac{f'(r)R'(r)}{4 f(r)R(r)} +\frac{R'^{2}(r)}{4R^{2}(r)} - \frac{f''(r)}{4f(r)} \bigg) \frac{\alpha(r)}{g(r)}}{+ \frac{f'(r)}{2 f(r) g(r)} \alpha'(r) },  \\
	c(r)&=\frac{M_{Pl}^{2} R'(r)}{4 R(r)}\bigg( \frac{f'(r)}{f(r)}+ \frac{g'(r)}{g(r)} - \frac{R''(r)}{2R(r)} +\frac{R'^{2}(r)}{R^{2}(r)} \bigg) \nonumber \\
	&  {-\bigg( \frac{f'^{2}(r)}{2f^{2}(r)} + \frac{f'(r)g'(r)}{4 g^{3/2}(r) \sqrt{f(r)}} - \frac{f'(r)R'(r)}{4 f(r)R(r)} +\frac{R'^{2}(r)}{4R^{2}(r)} - \frac{f''(r)}{4f(r)} \bigg) \alpha(r)} {+ \frac{f'(r)}{4f(r)} \alpha'(r). } \label{eqcr}  
	\end{align}
	
	{Equation for $\alpha (r)$ satisfies the following constraint equation,}
	\begin{align} \label{eqaaa}
	&{ \bigg( \frac{f'(r)}{f(r)} -\frac{R'(r)}{R(r)}\bigg) \alpha'(r)+ \bigg( \frac{f''(r)}{f(r)} -\frac{f'^{2}(r)}{2 f^{2}(r)} -\frac{R''(r)}{R(r)} -\frac{f'(r)g'(r)}{2 g^{3/2}(r) \sqrt{f(r)}} +\frac{f'(r)R'(r)}{2 f(r)R(r)} } \nonumber \\ 
	& {+\frac{g'(r)R'(r)}{2 g(r)R(r)}  + 2\frac{g(r)}{ R(r)} \bigg) \big(M_{Pl}^{2} + \alpha(r)\big) + \bigg(  \frac{R'^{2}(r)}{R^{2}(r)}  -\frac{f'^{2}(r)}{  f^{2}(r)}  - 2\frac{g(r)}{ R(r)} \bigg) \alpha(r) = 0}.
	\end{align}
For a given black hole background, $\alpha(r)$ can take different solutions, which essentially encodes the information about the hair and the underlying gravity theory. For simplicity we will take the trivial solution of the Eq.\ref{eqaaa} i.e $\alpha(r) =0$, provided we satisfy the following constraint:
	\begin{align} \label{constraint}
	{ \bigg( \frac{f''(r)}{f(r)} -\frac{f'^{2}(r)}{2 f^{2}(r)} -\frac{R''(r)}{R(r)} -\frac{f'(r)g'(r)}{2 g^{3/2}(r) \sqrt{f(r)}} +\frac{f'(r)R'(r)}{2 f(r)R(r)}  +\frac{g'(r)R'(r)}{2 g(r)R(r)}  + 2\frac{g(r)}{ R(r)} \bigg) =0}.
	\end{align}
In this paper, we will be considering mainly Schwarzschild, Schwarzschild-AdS and Schwarzschild-dS background solutions. One can check that for those backgrounds, Eq.\ref{constraint} is automatically satisfied. We will consider the more general cases involving $\alpha$ elsewhere. Considering the above background effective theory parameters, we are now in a position to consider the fluctuation dynamics, which is the quantity one experimentally observes. By using the diffeomorphism symmetry, in the next section, we will identify the Goldstone boson mode and its dynamical equation in the aforementioned hairy black hole background.  
	

	\subsection{St\"{u}ckelberg mechanism: Goldstone boson mode}
	
	We have already discussed the extra scalar degree of freedom is contained in the $r$-diffeomorphism broken Lagrangian given by Eq.\ref{eq2}. {As emphasised before, in the last section, we have expressed the $r$-symmetry broken Lagrangian (Eq.\ref{a}) in unitary gauge. Therefore, the associated Goldstone mode is absorbed inside the gravitational degrees of freedom. The well known St\"{u}ckelberg mechanism is all about extracting the aforesaid Goldstone mode by unfolding the unitary gauge. By this, one also can restore the $r$-symmetry which will be non-linearly realised by the mode. In order to restore the symmetry, one reparametrises again the broke symmetry as:
		\begin{align}
		r \rightarrow \tilde{r}&=r+\pi(x^{\mu}), \\ 
		t \rightarrow \tilde{t}&=t  \,\,,\, x^{i}\rightarrow \tilde{x}^{i}=x^{i},
		\end{align}
which will transform the background scalar field $\phi_0(r)$ as
\bea
\phi_0(\tilde{r}) = \phi_0(r) + \phi_0'(r) \pi .
\eea  
Therefore, by using this trick we can restore the  $r$-diffeomorphsim symmetry by considering the following combined transformations:
\bea
r\rightarrow r + \xi^{r}~~;~~\pi(x)\rightarrow \pi(x) - \xi^r ,
\eea
and simultaneously extract the dynamical degrees of freedom as $\pi(x)$ field which is identified as the Goldstone boson mode.} 
	Using these new coordinates, we separate out the Goldstone boson mode considering all transformed quantities:
	\begin{align}
	\tilde{g}^{\alpha\beta}&=g^{\alpha\beta}+\delta^{\beta}_{r}g^{\alpha\nu}\partial_{\nu}\pi+\delta^{\alpha}_{r}g^{\mu\beta}\partial_{\mu}\pi+\delta_{r}^{\alpha}\delta_{r}^{\beta}g^{\mu\nu} \partial_{\mu}\pi \partial_{\nu}\pi, \\
	\tilde{g}_{\mu\nu}&=g_{\mu\nu}-g_{r\nu}\partial_{\mu}\pi-g_{\mu r}\partial_{\nu}\pi+g_{rr}\partial_{\mu}\pi \partial_{\nu}\pi, \\
	{\tilde \partial}_{r}& =    \big(-\partial_{r}\pi +(\partial_{r}\pi)^{2}\big)\partial_{r} +{\partial}_{r},\\
 {\tilde \partial}_{\bar{i}}  &=  (-\partial_{\bar{i}} \pi + \partial_{\bar{i}}(\pi \partial_{r}\pi)) \partial_{r} + \partial_{\bar{i}}, \\
 {\delta \tilde{K}_{\bar{i}\bar{j}} } & {= - \bar{K}'_{\bar{i}\bar{j}} \pi + \frac{1}{\sqrt{g^{rr}}} \nabla_{\bar{i}}\partial_{\bar{j}} \pi}.
	\end{align}
	
	%
	%
	%
	%
	%
While calculating the action, in the present paper we will consider terms up to quadratic order in the $\pi$ field.
At this level, the tensor mode will not be coupled with the scalar mode. Therefore, the tensor mode analysis will be done separately. However, the reader can see the reference \cite{Franciolini:2018uyq} for details on this issue. Higher order interaction terms will be considered separately. Specifically, it would be interesting to understand the correction to the quasinormal modes taking into account loop effect. Finally the covariant quadratic Lagrangian for the Goldstone boson turns out to be,
\begin{align} \label{qww}
	\mathcal{L}_{\pi\pi}&=\frac{1}{4} c(r) g^{\mu\nu} \partial_{\mu}\pi \partial_{\nu}\pi -\frac{1}{2}  \bigg(\Lambda''(r)-\frac{c''(r)}{4g(r)} + {\frac{\alpha''(r)}{4g(r)}  g^{\bar{i}\bar{l}} g^{\bar{k}\bar{j}}  \partial_{r}g_{\bar{l}\bar{k}}
 \partial_{r}g_{\bar{i}\bar{j}} }\bigg)\pi^{2}+ \frac{c'(r)}{2g(r)}\pi \partial_{r} \pi   \nonumber \\
	&+\frac{2M_{2}(r)^{4}}{g^{2}(r)}(\partial_{r}\pi)^{2} {-\frac{\alpha(r)}{4} g^{\bar{i}\bar{m}} g^{\bar{n}\bar{j}}  \partial_{r}g_{\bar{m}\bar{n}} \bigg[ - \frac{ 1}{g(r)} \partial_{r}\left(g(r) \partial_{\bar{i}}\pi\partial_{\bar{j}}\pi\right) + \frac{g^{\alpha\beta}}{2}\partial_{r}g_{\bar{i}\bar{j}} \partial_{\alpha}\pi\partial_{\beta}\pi }\nonumber \\
& {+ g^{\bar{l}\bar{k}} \partial_{\bar{k}}\pi(\partial_{\bar{i}}\pi \partial_{r}g_{\bar{l}\bar{j}} + \partial_{\bar{j}}\pi \partial_{r}g_{\bar{i}\bar{l}} -\partial_{\bar{l}}\pi  \partial_{r}g_{\bar{i}\bar{j}}) \bigg]} 
{ - \frac{\alpha'(r)}{2}  g^{\bar{i}\bar{l}} g^{\bar{k}\bar{j}}  \partial_{r}g_{\bar{l}\bar{k}} \pi \nabla_{\bar{i}}\partial_{\bar{j}} \pi }  \nonumber \\
	&-\frac{1}{2} \bar{M}_{2}(r)^{2}g(r) \big(\bar{\Box} \pi \big)^{2} -\frac{1}{2}  \bar{M}_{3}(r)^{2} g(r) g^{\bar{i}\bar{k}}g^{\bar{l}\bar{j}} \big[\nabla_{\bar{i}}\partial_{\bar{l}}\pi \big] \big[\nabla_{\bar{k}} \partial_{\bar{j}}\pi \big] - \frac{\bar{M}_{1}(r)^{3}}{\sqrt{g(r)}}\partial_{r}\pi (\bar{\Box}\pi) \nonumber \\
	& {-\frac{1}{2} \left( \bar{M}_{2}(r)^{2} \bar{K}'^{\bar{i}}\,_{\bar{i}}  \bar{K}'^{\bar{j}}\,_{\bar{j}} +\bar{M}_{3}(r)^{2} \bar{K}'^{\bar{i}}\,_{\bar{j}}  \bar{K}'^{\bar{j}}\,_{\bar{i}} \right) \pi^{2} + \bar{M}_{2}(r)^{2} \bar{K}'^{\bar{i}}\,_{\bar{i}}\sqrt{g(r)} \pi \bar{\Box} \pi }\nonumber \\
	& {+ \bar{M}_{3}(r)^{2} \bar{K}'^{\bar{i}}\,_{\bar{j}} \sqrt{g(r) } \pi \nabla^{\bar{j}}\partial_{\bar{i}} \pi  + \bar{M}_{1}(r)^{3}  \frac{\bar{K}'^{\bar{i}}\,_{\bar{i}} }{g(r)} \pi \pi' - \bar{M}_{4}(r)^{2}  \frac{\bar{K}_{\bar{i}\bar{j}} \bar{K}'^{\bar{i}\bar{j}}  }{g(r)} \pi \pi' -2 M_{6}^{2}(r) \bar{K}'^{\bar{i}}\,_{\bar{i}} \partial_{r}  \bigg(\frac{\partial_{r} \pi}{g(r)} \bigg )  \pi } \nonumber\\
	& {+\frac{M_{11}(r)}{2} \partial_{r} g_{\bar{l} \bar{k}} \left( \bar{K}'^{\bar{i}}\,_{\bar{i}}  \bar{K}'^{\bar{l}\bar{k}} \pi^{2} - \bar{K}'^{\bar{i}}\,_{\bar{i}} \sqrt{g(r)} \pi \nabla^{\bar{l}}\partial^{\bar{k}} \pi - \bar{K}'^{\bar{l}\bar{k}} \sqrt{g(r)} \pi \bar{\Box} \pi \right)}\nonumber \\ 
	& {+\frac{M_{12}(r)}{2} \partial_{r} g_{\bar{i} \bar{l}} \left( \bar{K}'^{\bar{l}\bar{k}}  \bar{K}'^{\bar{i}}\,_{\bar{k}} \pi^{2} - 2 \bar{K}'^{\bar{l}\bar{k}} \sqrt{g(r)} \pi \nabla^{\bar{i}}\partial_{\bar{k}} \pi\right)}\nonumber \\ 
	&{+\frac{\lambda_{1}(r)}{4} \partial_{r} g_{\bar{i} \bar{l}} \partial_{r} g_{\bar{m} \bar{j}}g^{\bar{l}\bar{m}} \left( \bar{K}'^{\bar{k}}\,_{\bar{k}}  \bar{K}'^{\bar{i}\bar{j}} \pi^{2} - \bar{K}'^{\bar{k}}\,_{\bar{k}} \sqrt{g(r)} \pi \nabla^{\bar{i}}\partial^{\bar{j}} \pi - \bar{K}'^{\bar{i}\bar{j}} \sqrt{g(r)} \pi \bar{\Box} \pi \right)}\nonumber \\ 
	&+\frac{\hat{m}^{2}_{2}(r)}{2 g^{5}(r)}g'(r)^{2} (\partial_{r}\pi)^{2}+\frac{\hat{m}^{2}_{2}(r)}{2g^{3}(r)} (\partial_{r}^{2}\pi)^{2} - \frac{\hat{m}^{2}_{2}(r)}{g^{4}(r)}g'(r) \partial_{r}\pi \partial_{r}^{2}\pi  +  \frac{\hat{m}^{2}_{2}(r)}{2g^{2}(r)} g^{\bar{i}\bar{j}} (\partial_{r}\partial_{\bar{i}}\pi) (\partial_{r}\partial_{\bar{j}}\pi)\nonumber \\
	&  - \bar{M}_{4}^{2}(r) g^{\bar{i}\bar{l}} g^{\bar{k}\bar{j}}  \partial_{r}g_{\bar{l}\bar{k}}  \partial_{r} \pi \nabla_{\bar{i}}\partial_{\bar{j}} \pi + \frac{\lambda_{1}(r)}{4} \partial_{r} g_{\bar{i}\bar{j}} \partial_{r} g_{\bar{n}\bar{l}} g^{\bar{j}\bar{n}}g^{\bar{i}\bar{o}}g^{\bar{l}\bar{s}} \bar{\Box} \pi \nabla_{\bar{o}}\partial_{\bar{s}} \pi  \nonumber \\ 
    &{+ 4 M_{5}^{2}(r) \bigg[\partial_{r} \bigg(\frac{\partial_{r} \pi}{g(r)} \bigg ) \bigg]^{2}+ 2 M_{6}^{2}(r)\sqrt{g(r)} \partial_{r}  \bigg(\frac{\partial_{r} \pi}{g(r)} \bigg )  \bar{\Box}\pi - M_{7}(r) g^{\bar{i}\bar{l}} g^{\bar{k}\bar{j}}  \partial_{r}g_{\bar{l}\bar{k}} \partial_{r}  \bigg(\frac{\partial_{r} \pi}{g(r)} \bigg ) \nabla_{\bar{i}}\partial_{\bar{j}} \pi }\nonumber \\ 
&{+  \frac{M_{11}(r)}{2}\sqrt{g(r)} \partial_{r} g_{\bar{i}\bar{j}} g^{\bar{i}\bar{m}}g^{\bar{j}\bar{n}}\bar{\Box}\pi \nabla_{\bar{m}}\partial_{\bar{n}} \pi +  \frac{M_{12}(r)}{2}\sqrt{g(r)} \partial_{r} g_{\bar{i}\bar{j}} g^{\bar{i}\bar{n}}g^{\bar{l}\bar{k}}g^{\bar{j}\bar{m}} \nabla_{\bar{n}}\partial_{\bar{k}} \pi \nabla_{\bar{m}}\partial_{\bar{l}} \pi }\nonumber \\
&-\bigg({\frac{\hat{m}^{2}_{1}(r)}{ g(r)} g^{\bar{i}\bar{j}}  + M_{14}(r) (g^{rr})^{3/2} g^{\bar{a}\bar{i}} g^{\bar{b}\bar{j}} \partial_{r}g_{\bar{a}\bar{b}}  \bigg) \partial_{r}  \pi  \Bigg[-\partial_{\bar{l}} \pi \partial_{r}\Gamma^{\bar{l}}_{\bar{i}\bar{j}}+\partial_{\bar{j}} \pi \partial_{r} \Gamma^{\bar{l}}_{\bar{i}\bar{l}} - \partial_{\bar{l}}\big( g^{\bar{l}\bar{k}}  \partial_{\bar{i}}\pi \partial_{r}g_{\bar{k}\bar{j}}   \big)  }\nonumber \\ 
    &{+\partial_{\bar{j}}\bigg( \frac{g^{\bar{l}\bar{k}}}{2} \big( \partial_{\bar{i}}\pi \partial_{r}g_{\bar{k}\bar{l}} + \partial_{\bar{l}}\pi \partial_{r}g_{\bar{i}\bar{k}} \big) \bigg)- g^{\bar{m}\bar{k}} \big( \partial_{\bar{i}}\pi \partial_{r}g_{\bar{k}\bar{j}}  \big) \Gamma^{\bar{l}}_{\bar{l}\bar{m}} - \frac{g^{\bar{l}\bar{k}}}{2} \big( \partial_{\bar{l}}\pi \partial_{r}g_{\bar{k}\bar{m}} + \partial_{\bar{m}}\pi \partial_{r}g_{\bar{l}\bar{k}} \big) \Gamma^{\bar{m}}_{\bar{i}\bar{j}} }\nonumber \\
    &{ +  \frac{g^{\bar{m}\bar{k}}}{2} \big( \partial_{\bar{i}}\pi \partial_{r}g_{\bar{k}\bar{l}} + \partial_{\bar{l}}\pi \partial_{r}g_{\bar{i}\bar{k}} \big) \Gamma^{\bar{l}}_{\bar{j}\bar{m}} + \frac{g^{\bar{l}\bar{k}}}{2} \big( \partial_{\bar{j}}\pi \partial_{r}g_{\bar{k}\bar{m}} + \partial_{\bar{m}}\pi \partial_{r}g_{\bar{j}\bar{k}} \big) \Gamma^{\bar{m}}_{\bar{i}\bar{l}}\Bigg] },  
        	\end{align}
	

	where $\bar{\Box}=g^{\bar{i}\bar{j}}\nabla_{\bar{i}}\partial_{\bar{j}}$ and $\bar{i},\bar{j}=0,2,3$.  In the above Lagrangian, there exists higher time derivative terms such as $(\ddot{\pi}^{2}, \dddot{\pi}^2, \dot{\pi}\dddot{\pi})$. In general, those terms may contribute to unwanted ghosts, leading to instability, unless we fine tune the model parameters in such a way that at the equation of motion level it cancels all the higher derivative terms. For simplicity of our calculation, in order to remove the ghost we impose the following straightforward constraints on our effective theory parameters: $\bar{M}_{2}(r)^{2}=-\bar{M}_{3}(r)^{2}$ and $M_{11}(r)=- M_{12}(r)$ .   

	In the present work, as mentioned, we will consider simple cases where the background space time is asymptotically flat and de Sitter with scalar hair from the effective theory perspective. From the action, we can clearly see that the leading order kinetic term for the scalar field fluctuation is coming from the background $c(r)$ function. However, an important observation which will be discussed in detail in the subsequent section is that the $c(r)$ function becomes zero in the asymptotic region for asymptotically flat/dS/AdS black holes. Therefore, fluctuating Goldstone boson mode seems to become strongly coupled as one goes towards the asymptotic region of the black holes under consideration. This fact may have some interesting connection with the no-hair theorem, which we plan to study in the future. For the case of asymptotically AdS black holes, a hairy solution with the minimally coupled scalar field was first found for 2+1 dimensional black hole in \cite{3DHair} and in 2004 hairy solution for 3+1 dimensional black hole was found in  \cite{MTZ}. In both of these cases the well known Breitenlohner-Freedman (BF) bound \cite{BFbound} is satisfied and this guarantees that global AdS spacetime is stable under perturbations. While in the case of  \cite{gubser}, one needs to introduce local instability, specifically near the horizon either by introducing electromagnetic coupling or considering the mass of the real scalar field violating Breitenlohner-Freedman (BF) bound. For our present purpose, we will not be considering those situations. However, we can realise an interesting special case when the function $c(r)$ is zero throughout the region of the black hole space-time. Our Lagrangian seems to suggest that we still can have Goldstone boson fluctuations, whose dynamics will be controlled by the set of effective theory parameters $(M_i,\bar{M}_i,\hat{m}_i) $. Therefore, pure flat/dS/AdS metric can support the hair having the same symmetry as that of the black holes with well-behaved fluctuation. However, for these cases, the cosmological constant will be generated dynamically from the shift symmetry breaking scalar hair solution. In the subsequent sections, we will study in detail two special cases.

	\subsection{dS, AdS and flat Schwarzschild limit}	
	
	Our goal of this section would be to understand the details about the special cases we mentioned at the end of the previous section. Given the second order action for the Goldstone mode, it will be difficult to understand the fundamental properties of Goldstone mode. We therefore, restrict our discussion to very specific black hole backgrounds which are widely studied in the literature. Our major discussion will mainly be focused on asymptotically dS/flat Schwarzschild black holes. However, AdS case will be briefly discussed for completeness. We consider the following background metric components, 		
	\begin{eqnarray}
	f(r)&=1-\frac{2M}{r}+\epsilon \frac{r^{2}}{l_{c}^{2}},  \\
	g(r)&=\bigg(1-\frac{2M}{r}+\epsilon  \frac{r^{2}}{l_{c}^{2}}\bigg)^{-1},  \\
	R(r)&=r^{2} ~~;~~\phi=\phi^{\epsilon}_0(r),
	\end{eqnarray}
where, $\epsilon=(-1,1,0)$ correspond to asymptotically AdS, dS and flat Schwarzschild black holes respectively. For these metric functions, the background parameters of our effective field theory Eq.\ref{eq4} and Eq.\ref{eqcr} turns out to be $ \Lambda(r)=-\epsilon \frac{3 }{l_{c}^{2}} $, $c(r)=0$ {and $\alpha(r) = 0$}. Important to notice that the background cosmological constant is induced dynamically from the background scalar field. Any bare value of the cosmological constant can therefore be absorbed in the scalar field vacuum expectation value and the effective cosmological constant would be $\Lambda(r)$, which is constant.
	
In this limit the quadratic action for the scalar fluctuation boils down to the following expression:
	\begin{align} \label{qww}
	\mathcal{L}_{\pi\pi}&=\frac{2M_{2}(r)^{4}}{g^{2}(r)}(\partial_{r}\pi)^{2}
	+\frac{1}{2}\bar{M}_{3}(r)^{2}g(r) \big(\bar{\Box} \pi \big)^{2} -\frac{1}{2}  \bar{M}_{3}(r)^{2} g(r) g^{\bar{i}\bar{k}}g^{\bar{l}\bar{j}} \big[\nabla_{\bar{i}}\partial_{\bar{l}}\pi \big] \big[\nabla_{\bar{k}} \partial_{\bar{j}}\pi \big] \nonumber \\ 
	&- \frac{\bar{M}_{1}(r)^{3}}{\sqrt{g(r)}}\partial_{r}\pi (\bar{\Box}\pi) 
	+\frac{\hat{m}^{2}_{2}(r)}{2 g^{5}(r)}g'(r)^{2} (\partial_{r}\pi)^{2}+\frac{\hat{m}^{2}_{2}(r)}{2g^{3}(r)} (\partial_{r}^{2}\pi)^{2} - \frac{\hat{m}^{2}_{2}(r)}{g^{4}(r)}g'(r) \partial_{r}\pi \partial_{r}^{2}\pi \nonumber \\ 
	& +  \frac{\hat{m}^{2}_{2}(r)}{2g^{2}(r)} g^{\bar{i}\bar{j}} (\partial_{r}\partial_{\bar{i}}\pi)(\partial_{r}\partial_{\bar{j}}\pi) { - \bar{M}_{4}^{2}(r) g^{\bar{i}\bar{l}} g^{\bar{k}\bar{j}}  \partial_{r}g_{\bar{l}\bar{k}}  \partial_{r} \pi \nabla_{\bar{i}}\partial_{\bar{j}} \pi + 4 M_{5}^{2}(r) \bigg[ \partial_{r}\bigg(\frac{ \partial_{r}  \pi}{g(r)}\bigg) \bigg]^{2}} \nonumber \\
    & {+ 2 M_{6}^{2}(r) \sqrt{g(r)} \partial_{r}\bigg(\frac{ \partial_{r}  \pi}{g(r)}\bigg) \bar{\Box}\pi - M_{7}(r) g^{\bar{i}\bar{l}} g^{\bar{k}\bar{j}}  \partial_{r}g_{\bar{l}\bar{k}} \partial_{r}\bigg(\frac{ \partial_{r}  \pi}{g(r)}\bigg) \nabla_{\bar{i}}\partial_{\bar{j}} \pi }\nonumber \\ 
& {- \frac{M_{12}(r)}{2}\sqrt{g(r)} \partial_{r} g_{\bar{i}\bar{j}} g^{\bar{i}\bar{m}}g^{\bar{j}\bar{n}}\bar{\Box}\pi \nabla_{\bar{m}}\partial_{\bar{n}} \pi +  \frac{M_{12}(r)}{2}\sqrt{g(r)} \partial_{r} g_{\bar{i}\bar{j}} g^{\bar{i}\bar{n}}g^{\bar{l}\bar{k}}g^{\bar{j}\bar{m}} \nabla_{\bar{n}}\partial_{\bar{k}} \pi \nabla_{\bar{m}}\partial_{\bar{l}} \pi }\nonumber \\
&-\bigg({\frac{\hat{m}^{2}_{1}(r)}{ g(r)} g^{\bar{i}\bar{j}} } +{ \frac{M_{14}(r)}{ g^{3/2} (r)} g^{\bar{a}\bar{i}} g^{\bar{b}\bar{j}} \partial_{r}g_{\bar{a}\bar{b}} \bigg) \partial_{r}  \pi  \Bigg[-\partial_{\bar{l}} \pi \partial_{r}\Gamma^{\bar{l}}_{\bar{i}\bar{j}}+\partial_{\bar{j}} \pi \partial_{r} \Gamma^{\bar{l}}_{\bar{i}\bar{l}} - \partial_{\bar{l}}\big( g^{\bar{l}\bar{k}}  \partial_{\bar{i}}\pi \partial_{r}g_{\bar{k}\bar{j}}   \big)  }\nonumber \\ 
    &{+\partial_{\bar{j}}\bigg( \frac{g^{\bar{l}\bar{k}}}{2} \big( \partial_{\bar{i}}\pi \partial_{r}g_{\bar{k}\bar{l}} + \partial_{\bar{l}}\pi \partial_{r}g_{\bar{i}\bar{k}} \big) \bigg)- g^{\bar{m}\bar{k}} \big( \partial_{\bar{i}}\pi \partial_{r}g_{\bar{k}\bar{j}}  \big) \Gamma^{\bar{l}}_{\bar{l}\bar{m}} - \frac{g^{\bar{l}\bar{k}}}{2} \big( \partial_{\bar{l}}\pi \partial_{r}g_{\bar{k}\bar{m}} + \partial_{\bar{m}}\pi \partial_{r}g_{\bar{l}\bar{k}} \big) \Gamma^{\bar{m}}_{\bar{i}\bar{j}} }\nonumber \\
    &{ +  \frac{g^{\bar{m}\bar{k}}}{2} \big( \partial_{\bar{i}}\pi \partial_{r}g_{\bar{k}\bar{l}} + \partial_{\bar{l}}\pi \partial_{r}g_{\bar{i}\bar{k}} \big) \Gamma^{\bar{l}}_{\bar{j}\bar{m}} + \frac{g^{\bar{l}\bar{k}}}{2} \big( \partial_{\bar{j}}\pi \partial_{r}g_{\bar{k}\bar{m}} + \partial_{\bar{m}}\pi \partial_{r}g_{\bar{j}\bar{k}} \big) \Gamma^{\bar{m}}_{\bar{i}\bar{l}}\Bigg] }\nonumber \\ 
    &{+ \frac{\lambda_{1}(r)}{4} \partial_{r} g_{\bar{i}\bar{j}} \partial_{r} g_{\bar{n}\bar{l}} g^{\bar{j}\bar{n}}g^{\bar{i}\bar{o}}g^{\bar{l}\bar{s}} \bar{\Box} \pi \nabla_{\bar{o}}\partial_{\bar{s}} \pi} {- \frac{\lambda_{1}(r)}{4} \partial_{r} g_{\bar{i} \bar{l}} \partial_{r} g_{\bar{m} \bar{j}}g^{\bar{l}\bar{m}} \left( \bar{K}'^{\bar{k}}\,_{\bar{k}} \sqrt{g(r)}  \nabla^{\bar{i}}\partial^{\bar{j}} \pi + \bar{K}'^{\bar{i}\bar{j}} \sqrt{g(r)}  \bar{\Box} \pi \right) \pi}  \nonumber \\
    & { + \bigg[\frac{\bar{M}_{3}(r)^{2}}{2} \left( \bar{K}'^{\bar{i}}\,_{\bar{i}}  \bar{K}'^{\bar{j}}\,_{\bar{j}} -  \bar{K}'^{\bar{i}}\,_{\bar{j}}  \bar{K}'^{\bar{j}}\,_{\bar{i}} \right) +\frac{M_{12}(r)}{2} \left(\partial_{r} g_{\bar{i} \bar{l}}  \bar{K}'^{\bar{l}\bar{k}}  \bar{K}'^{\bar{i}}\,_{\bar{k}} - \partial_{r} g_{\bar{l} \bar{k}} \bar{K}'^{\bar{i}}\,_{\bar{i}}  \bar{K}'^{\bar{l}\bar{k}} \right) }\nonumber \\
    & {+\frac{\lambda_{1}(r)}{4} \partial_{r} g_{\bar{i} \bar{l}} \partial_{r} g_{\bar{m} \bar{j}} g^{\bar{l}\bar{m}}  \bar{K}'^{\bar{k}}\,_{\bar{k}}  \bar{K}'^{\bar{i}\bar{j}}  \bigg] \pi^{2} } {  + \bar{M}_{1}(r)^{3}  \frac{\bar{K}'^{\bar{i}}\,_{\bar{i}} }{g(r)} \pi \pi' - \bar{M}_{3}(r)^{2} \bar{K}'^{\bar{i}}\,_{\bar{i}}\sqrt{g(r)} \pi \bar{\Box} \pi }\nonumber \\
	&{+ \bar{M}_{3}(r)^{2} \bar{K}'^{\bar{i}}\,_{\bar{j}} \sqrt{g(r) } \pi \nabla^{\bar{j}}\partial_{\bar{i}} \pi - \bar{M}_{4}(r)^{2}  \frac{\bar{K}_{\bar{i}\bar{j}} \bar{K}'^{\bar{i}\bar{j}}  }{g(r)} \pi \pi' -2 M_{6}^{2}(r) \bar{K}'^{\bar{i}}\,_{\bar{i}} \partial_{r}  \bigg(\frac{\partial_{r} \pi}{g(r)} \bigg )  \pi } \nonumber\\
	& {+ \frac{M_{12}(r)}{2} \partial_{r} g_{\bar{l} \bar{k}} \left(  \bar{K}'^{\bar{i}}\,_{\bar{i}} \sqrt{g(r)}  \nabla^{\bar{l}}\partial^{\bar{k}} \pi + \bar{K}'^{\bar{l}\bar{k}} \sqrt{g(r)}  \bar{\Box} \pi \right) \pi} {- M_{12}(r) \partial_{r} g_{\bar{i} \bar{l}}  \bar{K}'^{\bar{l}\bar{k}} \sqrt{g(r)} \pi \nabla^{\bar{i}}\partial_{\bar{k}} \pi }.
	\end{align}%
 Our plan is to understand the properties of hairs from the effective field theory perspective. So far the usual approach to figure out the hairy black hole solutions was from the background effective field theory. In the present approach, we consider the effective theory of fluctuations. Most importantly, our starting point is the existence of background scalar hair which enjoys the same symmetry as that of the black holes. Importantly, this was one of the main criteria of the black hole no-hair theorem. Our study is based on the idea of the effective theory of inflation. In the inflation model, the approximate shift symmetry plays a very important role in constraining the theory parameters, which can be assumed to be time-independent and study the properties of fluctuation. However, for present case we do not have special symmetry which can help us to understand the nature of the effective theory parameters ($M_2(r)^{4},{\bar M}_3(r)^{2},{\bar M}_{1}^{3}(r), \hat{m}_{2}^{2}(r), {\bar{M}_{4}^{2}(r), M_{5}^{2}(r),M_{6}^{2}(r),M_{7}(r),M_{12}(r),M_{14}(r), \lambda_{1}(r),\hat{m}_{1}^{2}(r)}$).  Form the structure of our action we can proceed to construct the theory for a background which we will do in the next section. It is clear from our action that for three different asymptotic limits we may have hair with its detectable fluctuating degrees of freedom identified as $\pi(x)$ field. In order to understand further, let us consider following decomposition of $\pi=e^{-i\omega t}S(r)Y_{lm}(\theta,\phi)$, and the equation of motion for $S(r)$ turns out to be
	\begin{align}
	& 4 \frac{d}{dr}\left( M_{2}(r)^{4}f^{2}(r)r^{2} \frac{dS(r)}{dr}  \right) {+ \hat{m}^{2}_{1}(r)  \bigg[\frac{f'(r)}{ f(r)}r^{2} \omega^{2}- \bigg(\frac{f'(r)}{2 }+\frac{f(r)}{r}\bigg)l(l+1) \bigg] S'(r)} \nonumber \\ 
	&-\frac{d}{dr}\left(\hat{m}_{2}^{2}(r)r^{2}f(r)\frac{dS(r)}{dr}  \right) \omega^{2} + \frac{d}{dr}\left( \hat{m}_{2}^{2}(r)f^{2}(r)\frac{dS(r)}{dr}\right)l(l+1)    \nonumber \\
	&-\frac{d^{2}}{dr^{2}} \left[r^{2} \hat{m}_{2}^{2}(r)f^{3}(r) \frac{d^{2} S(r)}{dr^{2}} + r^{2} \hat{m}_{2}^{2}(r) f'(r) f^{2}(r) \frac{d S(r)}{dr} \right] + V_{l}(r) S(r)  \nonumber \\ 
	&+\frac{d}{dr} \left[r^{2} \hat{m}_{2}^{2}(r) f(r)f'(r)^{2} \frac{d S(r)}{dr}+ r^{2} \hat{m}_{2}^{2}(r) f'(r) f^{2}(r) \frac{d^{2} S(r)}{dr^{2}} \right]  
	\nonumber \\
 & {+8 \frac{d}{dr}\bigg[r^{2} M_{5}^{2}(r)  f'(r) \frac{d}{dr}\big(S'(r)f(r)\big) \bigg] - 8 \frac{d^{2}}{dr^{2}} \bigg[r^{2}  M_{5}^{2}(r)f(r) \frac{d}{dr}\big(S'(r)f(r)\big) \bigg]} \nonumber \\
 & { -4 M_{6}^{2}(r) \frac{f'(r)}{f^{1/2}(r)} l(l+1) S'(r)   - 4 \frac{d}{dr}\bigg[ r^{2}   \frac{M_{6}^{2}(r)  \omega^{2}}{f^{1/2}(r)}\bigg] S'(r) }  { - 4    \frac{ M_{6}^{2}(r) r^{2}  \omega^{2}}{f^{1/2}(r)} S''(r) }\nonumber \\
 &  { + 4 \frac{d}{dr}\bigg[  M_{6}^{2}(r)f^{1/2}(r)   \bigg] l(l+1) S'(r)  + 2\frac{d}{dr}\bigg[  M_{7}(r) \frac{f'(r)}{f(r)} r^{2}  \omega^{2} \bigg]S'(r) + 2M_{7}(r) \frac{f'(r)}{f(r)}r^{2}\omega^{2}S''(r) } \nonumber \\ 
 &{- 4 \frac{d}{dr}\bigg( M_{7}(r) \frac{f(r)}{r}\bigg) l(l+1) S'(r)  -  \frac{4}{r} M_{7}(r) f(r)l(l+1)  S''(r) }{- 2 M_{6}^{2}(r) \bar{K}'^{\bar{i}}\,_{\bar{i}}r^{2} \partial_{r}  \bigg(S'(r) f(r) \bigg )} \nonumber \\
&{+ M_{14}(r) f^{3/2}(r) \bigg[\frac{2 f'(r)}{f^{2}(r)} r \omega^{2}- \bigg(\frac{f'(r)}{r f(r)}+\frac{2}{r^{2}}\bigg)l(l+1) \bigg] S'(r)}\nonumber \\ 
& { + \bigg[ \bar{M}_{3}(r)^{2} \left( \bar{K}'^{\bar{i}}\,_{\bar{i}}  \bar{K}'^{\bar{j}}\,_{\bar{j}} -  \bar{K}'^{\bar{i}}\,_{\bar{j}}  \bar{K}'^{\bar{j}}\,_{\bar{i}} \right) + M_{12}(r) \left(\partial_{r} g_{\bar{i} \bar{l}}  \bar{K}'^{\bar{l}\bar{k}}  \bar{K}'^{\bar{i}}\,_{\bar{k}} - \partial_{r} g_{\bar{l} \bar{k}} \bar{K}'^{\bar{i}}\,_{\bar{i}}  \bar{K}'^{\bar{l}\bar{k}} \right) }\nonumber \\
    &{+\frac{\lambda_{1}(r)}{2} \partial_{r} g_{\bar{i} \bar{l}} \partial_{r} g_{\bar{m} \bar{j}} g^{\bar{l}\bar{m}}  \bar{K}'^{\bar{k}}\,_{\bar{k}}  \bar{K}'^{\bar{i}\bar{j}}  \bigg] r^{2} S(r) } 
     {   - \frac{d}{dr}\left(  \bar{M}_{1}(r)^{3} r^{2}  \bar{K}'^{\bar{i}}\,_{\bar{i}} f(r)+ \bar{M}_{4}(r)^{2} r^{2} \bar{K}_{\bar{i}\bar{j}} \bar{K}'^{\bar{i}\bar{j}}  f(r)\right) S(r)  } \nonumber \\
     &= \bigg[{\bar{M}_{3}(r)^{2} r^{2} \frac{\bar{K}'^{\bar{i}}\,_{\bar{i}}}{f^{3/2}(r)} + \frac{\lambda_{1}(r)}{4} \partial_{r} g_{\bar{i} \bar{l}} \partial_{r} g_{\bar{k} \bar{j}}g^{\bar{l}\bar{k}} r^{2} \frac{\bar{K}'^{\bar{i}\bar{j}}}{f^{3/2}(r) }} +\frac{d}{dr} \left(\frac{\bar{M}_{1}^{3}(r)r^{2}}{\sqrt{f(r)}}\right) {+ \frac{d}{dr}\bigg( \hat{m}^{2}_{1}(r)  r^{2}  \frac{f'(r)}{f(r)}\bigg)}\nonumber \\
	& {+ \frac{d}{dr}\bigg(r^{2} M_{4}^{2}(r) \frac{f'(r)}{f^{2}(r)} \bigg) -2 \frac{d}{dr}\bigg( r^{2} M_{6}^{2}(r) \frac{f'(r)}{f^{3/2}(r)}\bigg) } -  \frac{d}{dr}\bigg( r^{2}  M_{7}(r) \frac{f'^{2}(r)}{f^{2}(r)} \bigg)  
     -  \frac{d^{2}}{dr^{2}}\bigg( r^{2} M_{7}(r) \frac{f'(r)}{f(r)} \bigg) \nonumber \\
	&+ 2 \frac{d}{dr}\bigg( r M_{14}(r)   \frac{f'(r)}{f^{1/2}(r)}\bigg) \bigg]\omega^{2}S(r), 
	\end{align}
	where, $\omega$ is identified as  the frequency of the mode and  $l$ is the multipole number corresponding to the mode. $V_{l}(r)$ is called the ``effective potential" which takes the following form, 
	\begin{align}
	V_{l}(r)=&\Bigg[ { \bar{M}_{3}(r)^{2}  \frac{\bar{K}'^{\bar{i}}\,_{\bar{i}}}{f^{1/2}(r)} +\frac{\lambda_{1}(r)}{4} \partial_{r} g_{\bar{i} \bar{l}} \partial_{r} g_{\bar{k} \bar{j}}g^{\bar{l}\bar{k}} \frac{ \bar{K}'^{\bar{i}\bar{j}} }{f^{1/2}(r)}}- \frac{\bar{M}_{3}(r)^{2}}{r^{2}f(r)} \nonumber \\
	&- \frac{d}{dr} \left( \bar{M}_{1}^{3}(r) \sqrt{f(r)}\right)   {+ \frac{d}{dr}\bigg( \hat{m}^{2}_{1}(r)   \big(\frac{f'(r)}{2} +\frac{f(r)}{r}\big)\bigg)} { + \frac{\lambda_{1}(r)}{r^{4}} + \frac{d}{dr}\bigg( \frac{M_{4}^{2}(r)}{r} \bigg)  } \nonumber \\
 & {  - 2 \frac{d}{dr}\bigg( M_{6}^{2}(r) \frac{f'(r)}{f^{1/2}(r)}  \bigg)   + 2 \frac{d^{2}}{dr^{2}}\bigg(  M_{6}^{2}(r)f^{1/2}(r)  \bigg)  + 2 \frac{d}{dr}\bigg(  M_{7}(r)\frac{f'(r)}{r}\bigg)  } \nonumber \\
 &{- 2 \frac{d^{2}}{dr^{2}}\bigg( M_{7}(r) \frac{f(r)}{r}\bigg)} 
  {+ 2 \frac{M_{12}(r)}{r^{3}\sqrt{f(r)}}  +2 \frac{d}{dr}\bigg(M_{14}(r)  \frac{f^{3/2}(r)}{r} \big(\frac{f'(r)}{2f(r)} +\frac{1}{r}\big)\bigg)} \Bigg] l(l+1).
	\end{align}
One can clearly identify an important difference between the above fluctuation equation and the usual black hole linear perturbation equation for the scalar mode. The above structure of the effective equation seems to suggest that the underlying theory for the black hole scalar hair should be higher derivative in nature and therefore, in general, may not be stable under small fluctuation. Hence, it would be important to understand the underlying theory for the scalar hair from the effective theory perspective. This is what we are going to study in the next section. To this end let us emphasise the fact that for a given set of theory parameters we can in principle solve the above fluctuation equation and compute the quasinormal frequencies.{Therefore, as emphasised, we will restrict our study based on some assumptions which transform our complicated equation into simplified form. This assumptions and restrictions may be able to capture the qualitative nature of the quasi-normal frequencies, which we will discuss subsequently.} Considering all parameters to be zero except $M_{2}(r)$, $\bar{M}_{1}(r) $ and $\bar{M}_{3}(r)$, we get a simplified equation as 
	
	\begin{align}
	&4 \frac{d}{dr}\left( M_{2}(r)^{4}f^{2}(r)r^{2} \frac{dS(r)}{dr}  \right) + {V_{0}(r) S(r)} \nonumber \\
	&= \left[{\bar{M}_{3}(r)^{2} \left( r^{2} \frac{ f''(r)}{2 f^{2}(r)} - r^{2} \frac{f'^{2}(r)}{4 f^{3}(r)} +r \frac{f'(r)}{f^{2}(r)}- \frac{2}{f(r)} \right)}+\frac{d}{dr} \left(\frac{\bar{M}_{1}^{3}(r)r^{2}}{\sqrt{f(r)}}\right) \right]\omega^{2}S(r) .
	\end{align}
	
	{The potential $V_{0}(r)$ is}
	
	\begin{align}
	 {V_{0}(r)}&{=  \bar{M}_{3}(r)^{2} \left(r\frac{f''(r)f'(r)}{f(r)} -\frac{2 f''(r)}{f^{1/2}(r)} - r \frac{f'^{3}(r)}{2 f^{2}(r)}  +\frac{3f'^{2}(r)}{2 f(r)} +\frac{2 f(r)}{r^{2}} - \frac{2f'(r)}{r} \right)} \nonumber \\
	 &{ - \frac{d}{dr}\left[  \bar{M}_{1}(r)^{3} \left( r^{2} \frac{f''(r)f^{1/2}(r)}{2} - r^{2}  \frac{ f'^{2}(r)}{4 f^{1/2}(r)}  + r f'(r)f^{1/2}(r) -2f^{3/2}(r) \right) \right] } \nonumber \\
	 &{+\left[ \bar{M}_{3}(r)^{2} \left(\frac{ f''(r)}{2 f(r)} -\frac{f'^{2}(r)}{4 f^{2}(r)} +\frac{f'(r)}{rf(r)}- \frac{2}{r^{2}}  - \frac{1}{r^{2}f(r)}\right) - \frac{d}{dr} \left( \bar{M}_{1}^{3}(r) \sqrt{f(r)}\right)  \right]l(l+1)}
	\end{align}
	If one introduces the following coordinate transformation,
	$dr^* = {dr}/({2 M_2(r)^4 f^2(r) r^2})$,
	then above equation transforms into
	\begin{equation} \label{RW}
	\frac{d^2 S}{dr*^2}+(\omega_{eff}^2- V_{eff}) S = 0,
	\end{equation}
	where, the effective potential and the effective frequency spectrum are expressed as
	\begin{align}
	\omega_{eff}^2 &= -M_{2}(r)^{4}f^{2}(r)r^{2}\left[{\bar{M}_{3}(r)^{2} \left( r^{2} \frac{ f''(r)}{2 f^{2}(r)} - r^{2} \frac{f'^{2}(r)}{4 f^{3}(r)} +r \frac{f'(r)}{f^{2}(r)}- \frac{2}{f(r)} \right)}+\frac{d}{dr} \left(\frac{\bar{M}_{1}^{3}(r)r^{2}}{\sqrt{f(r)}}\right) \right]\omega^{2}, \\ 
	 V_{eff} &={- M_{2}(r)^{4}f^{2}(r)r^{2} V_{0}(r)} .
	\end{align}
	Note the important difference between Eq.\ref{RW} and usual Regge-Wheeler type black hole perturbation equation. The frequency being effective position dependent frequency function $\omega_{eff}$ and the potential is having complicated parameter dependences. Therefore, an important constraint has been arrived at from the above equations, which is ${\bar M}_1 \neq 0$. We only need to specify two parameters to understand the dynamics of fluctuations. Considering the usual definition, the effective frequency function should be finite in all the region of the radial coordinate. By choosing $\bar M_3=0$  and selecting the other two parameters appropriately as 
	\bea
	M_2(r)^4 \sim -\frac{1}{f(r)^2r^2},~{\bar M}_1(r)^3 \sim \frac{\sqrt{f(r)}}{r}, 
\eea	
	which enables us to write $\omega_{eff}$ and $V_{eff}$ as
	\bea
	&&\omega_{eff}^2 \sim \omega^2, ~~{\rm{and}}~\\
	&&V_{eff} \sim   - \frac{d}{dr}\left[ \frac{\sqrt{f(r)}}{r} \left( r^{2} \frac{f''(r)f^{1/2}(r)}{2} - r^{2}  \frac{ f'^{2}(r)}{4 f^{1/2}(r)}  + r f'(r)f^{1/2}(r) -2f^{3/2}(r) \right) \right]\nonumber\\&&~~~~~~~~~~~~- \frac{d}{dr} \left( \bar{M}_{1}^{3}(r) \sqrt{f(r)}\right)  l(l+1)  \label{wkb}.
	\eea
	All equations are written up to their dimensional constant. For this parametrisation we also have $r=r^*$. One particularly should notice the important difference that the centrifugal term appears as a multiplicative factor in the expression for the effective potential as opposed to the potential for scalar perturbation in the usual black hole background. 
	Some comments are in order for the complicated structure of the effective potential and the choices of parameters. First of all, it is to be noted that the $\omega_{eff}$ in its most general form is dependent on position. This itself is a problem for solving the equations of perturbations. The particular choice of the parameters helps one to write the frequency independent of position and thereby enabling one to use standard methods of finding the quasinormal frequencies. On the other hand, it is also very difficult to work with the general form of the equation. Hence, the simplified equation above with the potential (\ref{wkb}) can give us a hint into the behaviour of the black hole towards small perturbations. We will, therefore, look at the particular potential given in Eq. \ref{wkb}, via WKB approach \cite{will} to gather information about the nature of the quasinormal frequencies.
	
There are a few reasons for using the WKB approach to find out quasinormal modes. It is well known that these modes being the late time responses of the black holes to any linear order perturbations have already been observed in the gravitational wave signal from the black hole-black hole or black hole-neutron star mergers in recent gravitational wave detections. Therefore, the accuracy of calculation of the quasinormal modes is one of the most important issues among many others, because these modes will give the information of black hole parameters as well as they will help in constraining possible gravitational theories via observations. It is to be mentioned here different numerical techniques already are present in the literature to find quasinormal modes up to the desired accuracy. However, as pointed out clearly in \cite{konoplya_new}, that although these techniques are mainly based on convergent procedures, the analysis is extremely non-trivial and are different for different spacetimes. One, therefore, needs different separate numerical procedures to study different black hole space-times depending on the nature of the master differential equation. WKB method, on the contrary, provides a unique procedure, which on one hand remains unchanged for various different master equations and on the other hand, it provides sufficient numerical accuracy too. Because our goal is to look at different toy models in various black hole backgrounds, we found the WKB method to be a universal tool to comment about the stability of the space-time.
	
	It has been already shown \cite{konoplya_new, konoplya1} that when one increases the order in the WKB series from three to six, the relative error diminishes drastically by a few times or even by orders. However, in many scenarios, this formula does not allow one to compute quasinormal modes which have $n\geq l$ with much accuracy. But, in this work, we have particularly omitted the modes with $n\geq l$ to get some idea about the behaviour of the perturbations with sixth order WKB method and keeping our main target as the understanding of hairy black holes in the effective field theory framework. 
	
	 The sixth order WKB method developed by Konoplya \cite{konoplya1} gives values accurately as one gets by performing numerical integration of the master differential equation. The sixth order WKB formula for a general black hole potential $V (r)$ is given by
	\begin{figure}[h] 
 	\begin{minipage}[c]{\textwidth}
 		\centering
 		\includegraphics[width=3in]{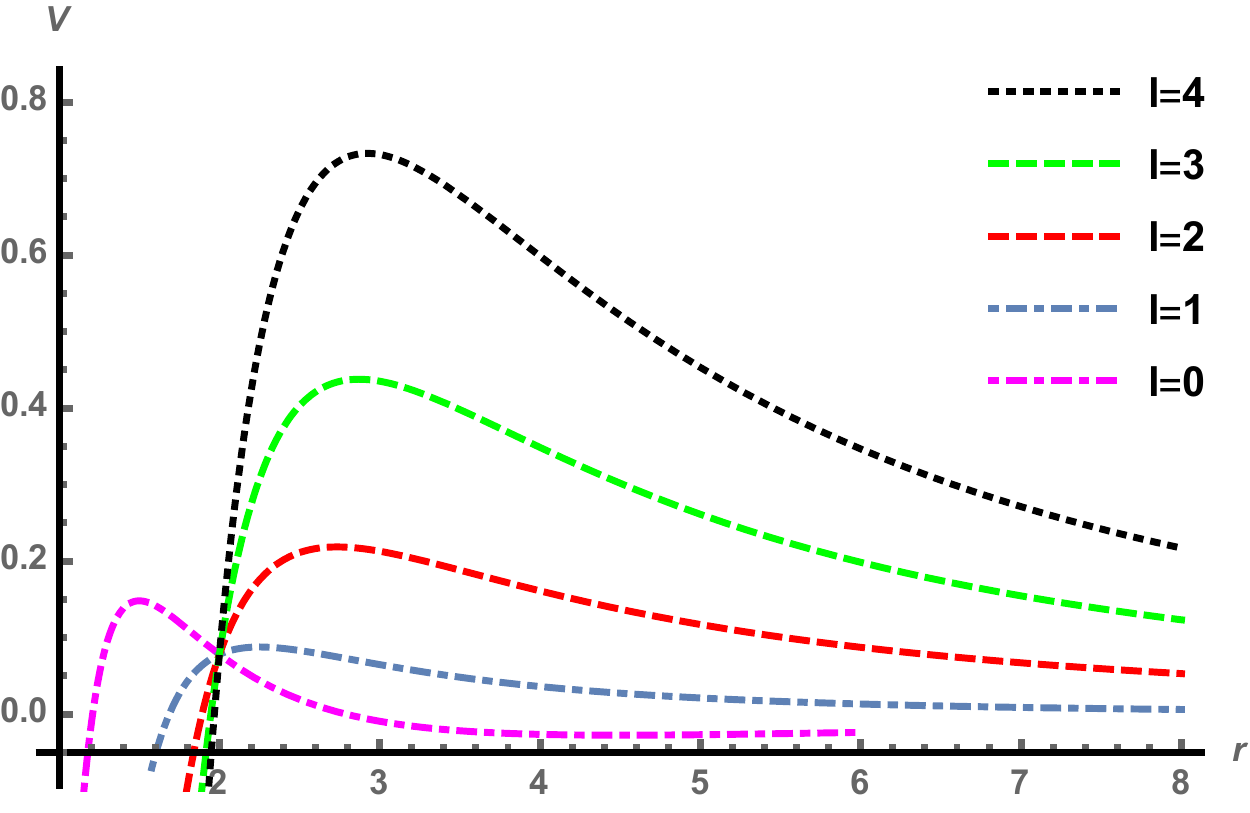}
 		\includegraphics[width=3in]{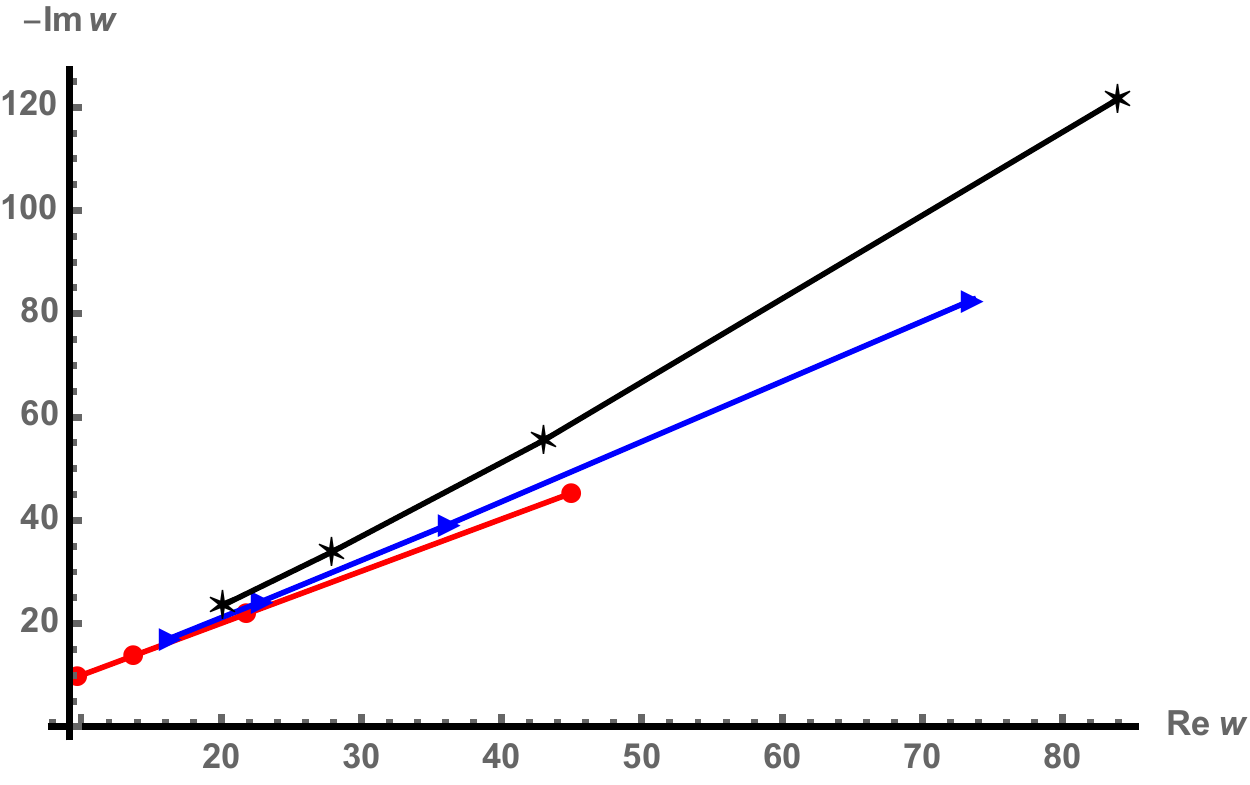}
 		\caption{Left one is a plot of the potential vs radial distance for different $l$ values with $n=0$. $l$ increases from $0$ to $4$ in steps of $1$ from bottom to top. Asymptotically flat Schwarzschild geometry is chosen as the metric function in the potential (\ref{wkb}). The right one is a plot of Re. vs. Im. $\omega$ for different values of $l$. The red curve has $n=0$, blue has $n=1$ and black has $n=2$, showing clear difference with the standard behaviour of QN frequencies with multipole number and overtone.}
 	\end{minipage}
 \end{figure}
\bea
	\frac{i(\omega^2-V(r_0))}{\sqrt{-2V^{''}(r_0)}}-\Lambda_2-\Lambda_3-\Lambda_4-\Lambda_5-\Lambda_6 = n+\frac{1}{2}.\label{qnmeqn}
	\eea
Here, $V(r_0)$ is peak value of $V(r)$ , $V^{''}(r_0) = \frac{d^2V}{dr_*^2}|_{r=r_0}$ , $r_0$ is the value of the radial coordinate corresponding to the maximum of the potential $V(r)$ and $n$ is the overtone number. $\Lambda_i$'s are the higher order WKB correction terms, whose expressions can be found in \cite{konoplya1}. Thus, using the potential $V_{eff}$ given by Eq.\ref{wkb} in Eq.\ref{qnmeqn}, we plot (see Fig. (1) and (2)) the nature of the potential (for arbitrary choices of the parameters), variation of the real and imaginary parts of the quasinormal frequencies with multiple number $l$ and overtone $n$. In finding out the quasinormal frequencies, we have used the asymptotically flat Schwarzschild metric as the metric function in the effective potential.		
	 \begin{figure}[h] 
 	\begin{minipage}[c]{\textwidth}
 		\centering
 		\includegraphics[width=3in]{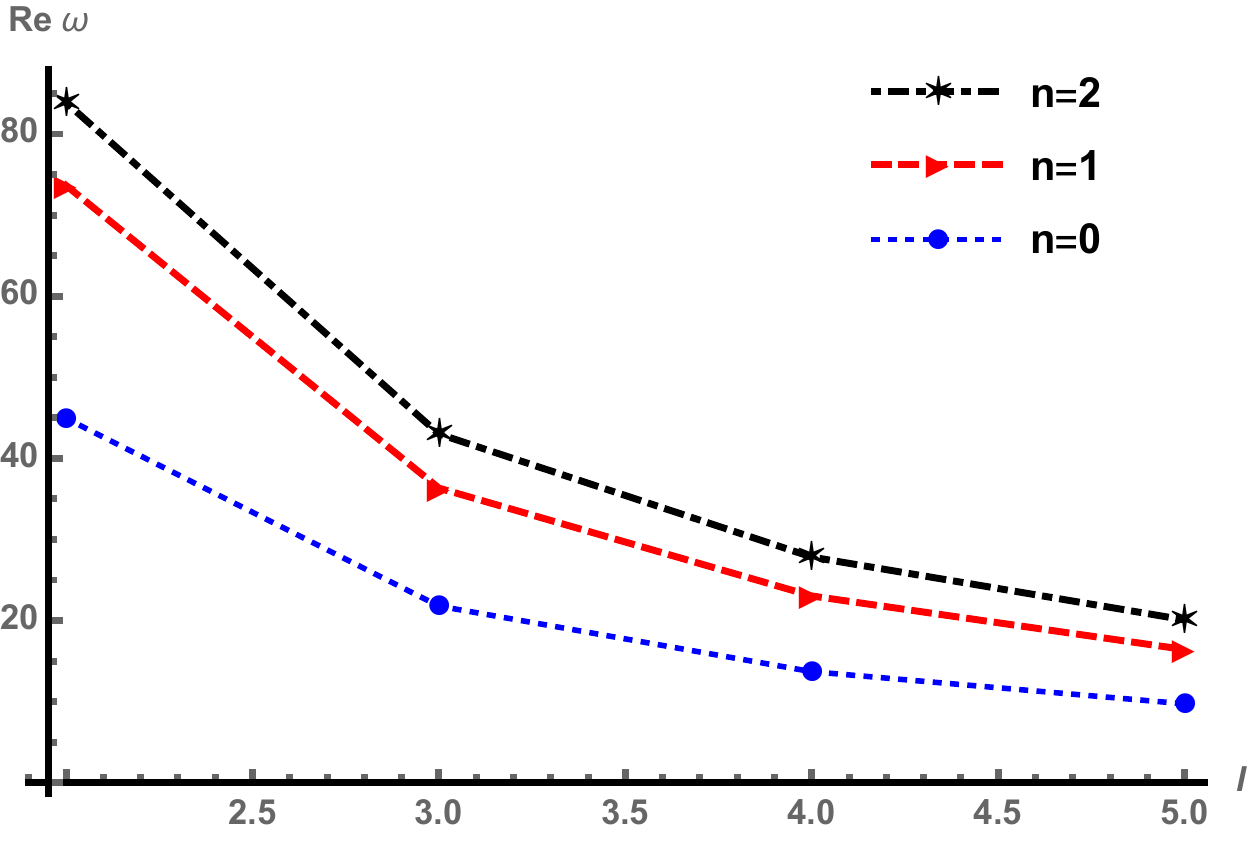}
 		\includegraphics[width=3in]{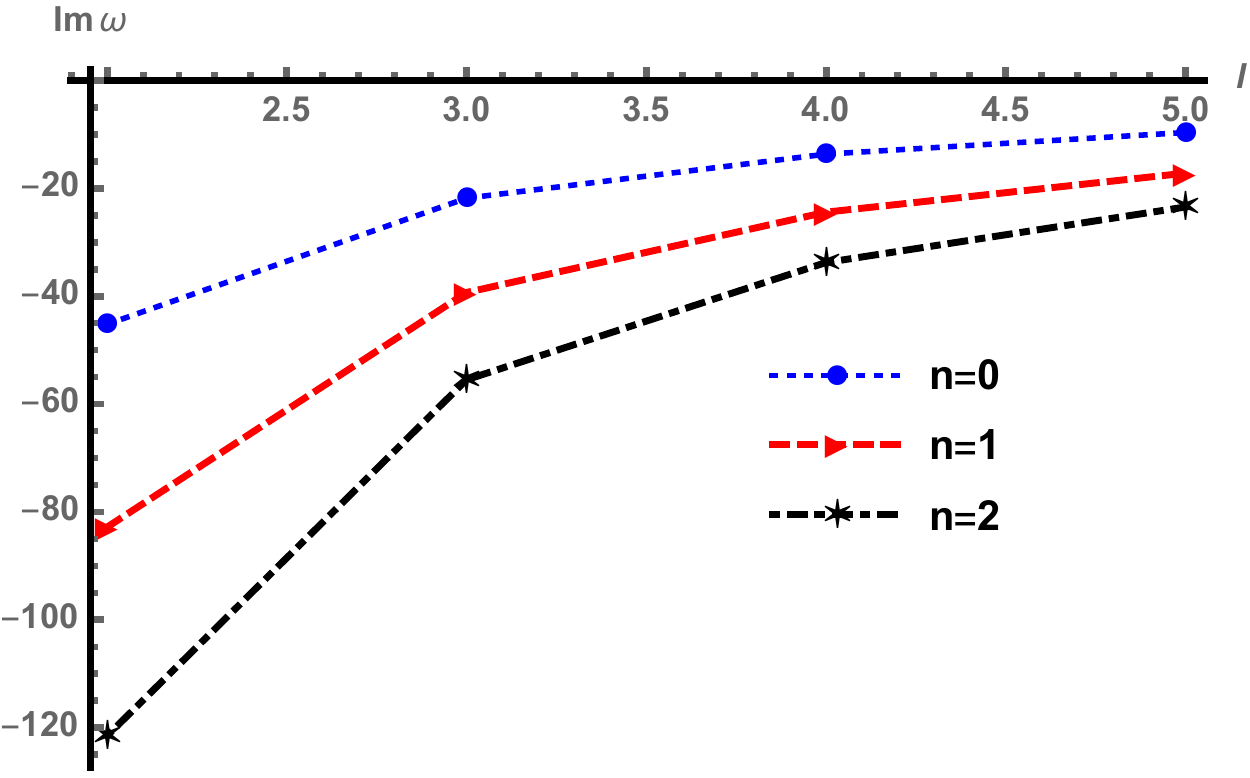}
 		\caption{Plot of the real and imaginary parts of $\omega$ vs multipole number $l= 2, 3, 4$ and $5$ for different $n$ values. The blue dashed plot corresponds to $n=0$, red dotted curve corresponds to $n=1$ and black one corresponds to $n=2$. }
 	\end{minipage}
 \end{figure}
	Note the fundamental difference of the behaviour of the quasinormal frequencies with changing the overtone ($n$) and multipole number ($l$). In general, for the four dimensional Schwarzschild black hole in asymptotically flat background, the real oscillation frequency decreases and imaginary part of the frequency increases with increase of the overtone number while the multipole number is kept fixed. However, in our case, both the real oscillation frequency as well as the imaginary part of the frequency increases with increase of the overtone. We will see in the next sections, that this behaviour remains for a specific theory which contains all the information discussed above. 

From the stability point of view, it is well known that  the evolution of the perturbations in a black hole spacetime has three distinct parts: firstly there is a response to perturbations at very early times, where the form of the signal depends crucially on the initial conditions, secondly at an intermediate stage, the signal is dominated by a quasinormal frequencies, i.e. the exponential decay, also called the ring down phase. The frequency and damping times are determined by the quasinormal modes. In this particular phase, the signal depends only on the black hole parameters and finally due to the backscattering of the curvature of the space-time, at late times the propagating wave shows a falloff of the field at the tail phase. This tail phase is completely independent of the initial data and it persists even if there is no horizon. In this work, we neglect the back reaction and solely focus on the second part of the perturbative response, i.e. the ringing phase. 
	
		For the general case, as we will see in the next section, for a particular underlying model, it is very difficult to identify the effective potential.  More general discussion with the unknown parameters will be done in a separate paper later. 
	
	%
	%
	%
	%
	%
	%
	%
	\section{Underlying theory}
	
	Based on our previous analysis, in this section, we will consider a possible underlying model which has a hairy solution with the flat/dS black holes. Our approach here will be the same as the model of ghost condensation \cite{ghostcondensate}. However, an important difference lies in the existence of spherically symmetric scalar hair. Therefore, the order parameter for spontaneous breaking of $r$-translation symmetry is given by vacuum expectation value of $\langle\partial_{\mu} \phi\rangle$. From this, it is supposed that the system has shift symmetry which also implies the absence of non-derivative coupling.
We are considering the spherically symmetric black holes with scalar hair which enjoys the same symmetry as the black hole metric. Hence, we have $\phi \equiv \phi(r)$, and therefore, $\langle\partial_{r}\phi\rangle \neq 0$. Given the effective action we derived in the previous section, we consider following two functions of the basic composite field variable $X=-1/2 (\partial \phi)^2$ as  $P(X)=-\alpha X^{2}+  \beta X^{4}$, and $F(X)=\tilde{M}_{3} g^{\mu\nu}\partial_{\mu}X \partial_{\nu}X $. For our purpose, we call $P(X)$ as kinetic potential term and $F(X)$ as kinetic gradient term. All the parameters, $\alpha, \beta$, are positive dimensionful constant with $\alpha = \tilde{M}_{1}^{-4}, \beta = \tilde{M}_{2}^{-12}$ and $\tilde{M}_{3} \equiv \tilde{M}_{3}^{-6}$. All $\tilde{M}$'s are of dimension one. The Lagrangian for the above scalar hair will be expressed as

	\begin{align} 
	\mathcal{L}_{\phi}&=\sqrt{-g}\left[P(X)+F(X)\right] \nonumber, \\ 
	&=\sqrt{-g}\left[-\tilde{M}_{1}^{-4} X^{2}+  \tilde{M}_{2}^{-12} X^{4}+\tilde{M}_{3}^{-6} g^{\mu\nu}\partial_{\mu}X \partial_{\nu}X  \right].
	\end{align}
	The expression for the action will be
	\begin{align} \label{action}
	\mathcal{S}&= \int d^{4}x \bigg[\sqrt{-g} \frac{M_{Pl}^{2}}{2} \mathcal{R} - \mathcal{L}_{\phi} + \mathcal{J} \bigg],
	\end{align}
	where, $\mathcal{J}$ is the parameter we will tune to get different values of the cosmological constant. Here, we will not discuss this fine tuning issue. Our goal of this paper would be to understand the quasinormal modes of those hairy black holes, where the condensed spherically symmetric scalar hair contributes to the effective cosmological constant which can add up with the bare cosmological constant parameterised by $\mathcal{J}$. The expression for the stress-energy tensor appears to be
	\begin{align} \label{9}
	T_{\mu\nu}&= -P(X) g_{\mu\nu}+\mathcal{J} g_{\mu\nu} - 2P'(X) \partial_{\mu} \phi \partial_{\nu} \phi - F(X) g_{\mu\nu} \nonumber \\ 
	& + 2 \tilde{M}_{3}^{-6} \partial_{\mu}X \partial_{\nu}X -4  \tilde{M}_{3}^{-6} g^{\alpha\beta}\partial_{\alpha}(\partial_{\mu} \phi \partial_{\nu} \phi) \partial_{\beta}X.
	\end{align}
	At the minimum of the kinetic potential, $P(X)=- \tilde{M}_{1}^{-4} X^{2}+ \tilde{M}_{2}^{-12} X^{4}$ ,
	\begin{align}
	X_{0}= \pm \sqrt{\frac{\tilde{M}_{2}^{12}}{2 \tilde{M}_{1}^{4}}}=\frac{\tilde{M}_{2}^{6}}{\sqrt{2}\tilde{M}_{1}^{2}} \equiv \pm c,
	\end{align}
	the energy momentum tensor takes the following form,
	\begin{align}
	T_{\mu\nu}&=- P(X_{0}) g_{\mu\nu} +\mathcal{J} g_{\mu\nu} .
	\end{align}
	Therefore, by appropriate choice of the value of ${\cal J}$, we can have spherically symmetric hairy black holes which are asymptotically dS/AdS/flat. The solution for the scalar field is therefore parametrised by 
	\begin{align}
	P'(X_{0})=0, \,\,\,\, &F(X_{0})=0, \,\,\,\, \partial_{\mu}X_{0}=0, \nonumber \\
	\Rightarrow \phi_{0}(r)&= c^{1/2} \int \frac{dr}{(f(r))^{1/2}}.
	\end{align}  
	At this point, it is worth pointing out that for $X_0 >0$ the solution must be of the cosmological type which is precisely the ghost condensation model \cite{ghostcondensate}. However, in order to have scalar hair with the same symmetry as that of the black hole, we need to consider $ {X_0<0}$. Considering the value of $X_0$, one arrives at $P(X_{0}=-c)=-\frac{\tilde{M}_{2}^{12}}{4 \tilde{M}_{1}^{8}} < 0$ which contributes negative cosmological constant (AdS) in the background. Therefore, by tuning the value of ${\cal J}$, we can also obtain dS and Schwarzschild solutions. Once we have spherically symmetric hair, the Lagrangian for the Goldstone fluctuation $\phi= \phi_0(r)+ \varphi(x)$, turns out to be  
	\begin{align}
	\mathcal{L}_{\varphi\varphi}=  &\bigg[4\left[4 \tilde{M}_{2}^{-12} (g^{rr})^{4} (\partial_{r}\phi_{0})^{6}+ \tilde{M}_{3}^{-6} g^{rr} [\partial_{r}(g^{rr}\partial_{r}\phi_{0})]^{2}\right] (\partial_{r}\varphi)^{2} +4\tilde{M}_{3}^{-6} g^{rr}(g^{rr}\partial_{r}\phi_{0})^{2} (\partial_{r}^{2}\varphi)^{2} \nonumber \\ 
	&+8 \tilde{M}_{3}^{-6} (g^{rr})^{2} \partial_{r}\phi_{0} \partial_{r}(g^{rr}\partial_{r}\phi_{0}) \partial_{r}\varphi \partial_{r}^{2}\varphi - 4\tilde{M}_{3}^{-6} g^{rr}(\partial_{r}\phi_{0})^{2} (\partial_{r}\dot{\varphi})^{2}  \nonumber \\
	&+4\tilde{M}_{3}^{-6} g^{ij}(g^{rr}\partial_{r}\phi_{0})^{2} (\partial_{r}\partial_{i}\varphi) (\partial_{r}\partial_{j}\varphi) + 16\tilde{M}_{2}^{-12} X_{0}^{2} g^{rr} \partial_{r}\phi_{0} (g^{\mu\nu}\partial_{\mu}\varphi \partial_{\nu}\varphi) \partial_{r} \varphi \nonumber \\
	&+4 \tilde{M}_{3}^{-6} g^{\mu\nu} \partial_{\mu} \big(g^{rr} \partial_{r}\phi_{0} \partial_{r}\varphi \big) \partial_{\nu} \big(g^{\alpha\beta}\partial_{\alpha}\varphi \partial_{\beta}\varphi \big)+32\tilde{M}_{2}^{-12} X_{0}^{2} (g^{rr})^{2} \partial_{r}\phi_{0} ( \partial_{r} \varphi)^{3} \bigg].
	\end{align}
	Now, let us mention the  connection between the scalar field fluctuation $\delta \phi = \varphi(x)$ with the previously discussed Goldstone boson mode in the gravity sector. Considering the general coordinate transformation $r\rightarrow r + \pi$, the scalar field transforms as
	\bea
	\phi'(x') = \phi_0(r+\pi) +\varphi(r+\pi) = \phi_0(r) + \varphi(x) + \pi(x) \phi'_0(r) +{\cal O}(\pi^2).
	\eea
	Hence, to the linear order the scalar field fluctuation $\varphi$ is identified with the Goldstone mode $\pi(x)$ as  
	\bea
	\delta \phi(r)\equiv \varphi = - \pi \phi_{0}'(r).
	\eea
	More importantly, in the spherically symmetric hairy background, the kinetic potential term $P(X)$ does not contribute to the time variation of the fluctuation. This is the reason we have introduced kinetic gradient term $F(X)$. Therefore, $X$ can be thought of as an effective composite degree of freedom which is behaving like a Higgs field and the formation of spherically symmetric hair is essential and is similar to the well known Higgs mechanism in terms of the composite field $X$ of dimension four operator. By combining the bare cosmological constant ${\cal J}$ and the vacuum of the scalar field hair, the net effective cosmological constant is expressed as 
	\begin{align}
	\Lambda= \frac{3}{l_{c}^{2}} = \frac{1}{M_p^2}\left(\frac{\tilde{M}_{2}^{12}}{4 \tilde{M}_{1}^{8}} + \cal{J}\right)  = \frac{1}{M_p^2}\left(\frac{X_{0}^{2}}{2\tilde{M}_{1}^{4}} + \cal{J}\right). 
	\label{001a}
	\end{align} 
	The dependence of cosmological constant on scalar hair is seen in the above equation.
	The governing equation for the radial component of the scalar field fluctuation $\varphi(x)=e^{-i\omega t}S(r)Y_{lm}(\theta,\phi)$
	will take the form of the following fourth order differential equation,
	\begin{align}  
	S^{''''}&(r) + 4 \left(\frac{f'(r)}{f(r)}+\frac{1}{r}\right) S'''(r) \nonumber \\
	-\bigg[ &\frac{2\tilde{M}_{3}^{6}}{\tilde{M}_{1}^{4}f(r)} -\frac{9f'^{2}(r)}{4f^{2}(r)}-\frac{5f''(r) }{2f(r)}- \frac{9f'(r)}{f(r)r}-\frac{2}{r^{2}}  - \frac{\omega^{2}}{f^{2}(r)} +\frac{l(l+1)}{f(r)r^{2}} \bigg] S''(r) \nonumber \\
	- \bigg[2 &\frac{\tilde{M}_{3}^{6}}{ \tilde{M}_{1}^{4}}  \left( \frac{f'(r)}{f^{2}(r)}+\frac{2}{f(r)r} \right)-\frac{f'(r)f''(r)}{f^{2}(r)} -\frac{3f'^{2}(r)}{2f^{2}(r)r} - \frac{f'''(r)}{2f(r)} \nonumber \\
	& -2 \frac{f''(r)}{f(r)r}-\frac{f'(r)}{f(r)r^{2}}- \frac{2 \omega^{2}}{f^{2}(r)r} +\frac{l(l+1)f'(r) }{f^{2}(r)r^{2}}\bigg] S'(r) =0.
	\end{align} 
	In order get the numerical solution for the quasinormal modes, it is convenient to express the above equation in terms of the dimensionless coordinate $\tilde{r}=\frac{r}{r_{0}}$, theory parameter $m_{0}^{2}=\frac{\tilde{M}_{3}^{6}r_{0}^{2}}{\tilde{M}_{1}^{4}}$ and $W_{0}^{2}=(r_{0}\omega)^{2}$. So the above equation in the new coordinate is
	\begin{align}  \label{eq29}
	&\frac{d^{4}S(\tilde{r})}{d\tilde{r}^{4}} + 4  \left(\frac{f'(\tilde{r})}{f(\tilde{r})}+\frac{1}{\tilde{r}}\right) \frac{d^{3}S(\tilde{r})}{d\tilde{r}^{3}} \nonumber \\
	-\bigg[ &\frac{2 m_{0}^{2}}{f(\tilde{r})} -\frac{9f'^{2}(\tilde{r})}{4f^{2}(\tilde{r})} -\frac{5f''(\tilde{r}) }{2f(\tilde{r})}- \frac{9f'(\tilde{r})}{f(\tilde{r})\tilde{r}}- \frac{2}{\tilde{r}^{2}} - \frac{W_{0}^{2}}{f^{2}(\tilde{r})} +\frac{l(l+1)}{f(\tilde{r})\tilde{r}^{2}} \bigg] \frac{d^{2}S(\tilde{r})}{d\tilde{r}^{2}} \nonumber \\
	- \bigg[2 & m_{0}^{2}  \left( \frac{f'(\tilde{r})}{f^{2}(\tilde{r})}+ \frac{2}{f(\tilde{r})\tilde{r}} \right) -\frac{f'(\tilde{r})f''(\tilde{r})}{f^{2}(\tilde{r})} -\frac{3f'^{2}(\tilde{r})}{2f^{2}(\tilde{r})\tilde{r}} - \frac{f'''(\tilde{r})}{2f(\tilde{r})} \nonumber \\
	& -2 \frac{f''(\tilde{r})}{f(\tilde{r})\tilde{r}}- \frac{f'(\tilde{r})}{f(\tilde{r})\tilde{r}^{2}}- \frac{2 W_{0}^{2}}{f^{2}(\tilde{r})\tilde{r}} +\frac{l(l+1)f'(\tilde{r}) }{f^{2}(\tilde{r})\tilde{r}^{2}}\bigg] \frac{dS(\tilde{r})}{d\tilde{r}} =0
	\end{align}
 This is the final master equation which we will be analysing for two different black hole backgrounds. Extra care must be taken as our master equation is fourth order in the derivative. We have two theory parameters $(m_0, X_0)$ which are functions of $(\tilde{M}_1,\tilde{M}_2,\tilde{M}_3)$. $X_0$ sets the background value of the cosmological constant and also provides a scalar field profile. Now, in the following two sub-sections, our goal would be to compute the quasinormal modes considering asymptotically flat Schwarzschild and Schwarzschild de Sitter background. The value of $X_0$ will be determined by the present value of the cosmological constant of our universe. For our analysis we consider two sample values of $m_{0}^{2}=10^{-1}$ and $10^{-6}$.
 \subsection{Quasinormal mode analysis in asymptotically dS/flat hairy black holes} 
In anti-de Sitter space, we already know the existence of spherically symmetric scalar hair \cite{MTZ,gubser} for minimally coupled and two derivative scalar field. However, for asymptotically dS/flat black hole background it has been proved to be very difficult to find a hairy solution where hair also enjoys the same symmetry as the black holes. However, if one relaxes these assumptions, the time-dependent scalar hair in the spherically symmetric black holes have already been found \cite{radu}. However, for a general shift symmetric scalar field theory known as Horndeski theory with an additional non-minimal coupling with Gauss-Bonnet gravity, can give rise to spherically symmetric scalar hair as has recently been shown in \cite{GBhair}. To the best of our knowledge, we are for the first time pointing out the existence of scalar hair with same symmetry as that of the black holes for a minimally coupled but higher derivative scalar field theory introduced earlier. 

For simplicity, we also consider the shift symmetric theory, which we found easy to construct, observing the effective theory Lagrangian derived in previous sections. Our goal in this subsection would be to understand the behaviour of quasinormal mode frequencies in the aforementioned hairy black hole background for asymptotically flat and dS black holes. AdS black holes will be discussed in a separate publication with their implications to AdS/CFT  in details. 
 
 Before going into the discussion of the quasinormal frequencies for black holes in the above-mentioned background, a little bit discussion on black hole perturbations and corresponding quasinormal modes are in order. A perturbed black hole is generally described by the metric $g_{\alpha\beta}=g^0_{\alpha\beta}+\delta g_{\alpha\beta}$, where $g^0_{\alpha\beta}$ is the background space time, i.e. the space time of the non-perturbed black hole and $\delta g_{\alpha\beta}$ denotes the perturbation over the background. In the linear approximation, the perturbations $\delta g_{\alpha\beta}$ are much less than the background, i.e. $\delta g_{\alpha\beta}<< g^0_{\alpha\beta}$. Once we perturb a black hole by some fields or by perturbing the metric itself, the background responds to the perturbations by the emission of gravitational waves which evolves in time in the following three stages: first, there is a relatively short period of initial outburst of radiation; second, a long period of damped oscillations dominated by the quasinormal modes and lastly, at very late times the quasinormal modes are suppressed by power law or exponential tails in certain specific space-time geometries (see \cite{Nollert:1999ji, Kokkotas:1999bd, Berti:2009kk, Konoplya:2011qq} for a review). For all practical purposes, the second stage, i.e. the quasinormal ringing is the most important one in the context of gravitational waves, because these modes carry unique information about the black hole parameters. 
 	
 	In general, in a spherically symmetric black hole background, the study of perturbations due to linearised fields (with spin 0, 1 or 2) can be reduced to the study of a Schr\"{o}dinger like second order differential equation. To determine the oscillation modes of a black hole, which corresponds to the solutions of the mentioned differential equation, one has to impose physically motivated boundary conditions at the two boundaries of the problem, viz. at the horizon and spatial infinity. For the spacetimes of our interest, the potential in the Schr\"{o}dinger like equation goes to zero at the horizon and at asymptotic infinity/cosmological horizon, and in this limit solutions to the wave equations are purely ingoing at the horizon and purely outgoing at infinity. This means that at the classical level, nothing should leave the horizon and nothing should come out from infinity to disturb the system.
 
  Quasinormal modes differ from other problems involving small oscillations in the sense that the black hole system is dissipative. Waves can move either to infinity or into the black hole horizon. Therefore, a normal mode analysis of the system is not possible. However, there exists a discrete infinity of quasinormal modes, which are defined as eigenfunctions of the operators describing the governing equation of the perturbation, which satisfies the above-mentioned boundary conditions. The corresponding eigenfrequencies consist of a real and an imaginary part, the latter is related to the damping time of the mode. Quasinormal frequencies are sorted out generally by their imaginary part and are labeled by an integer $n$, which, in the literature is called the overtone number. The fundamental quasinormal mode corresponding to the overtone number $n = 0$ is the least damped mode, and it usually dominates the ringdown waveform because it has the smallest imaginary part and therefore it is the longest-lived mode.
 
 We have already introduced the dimensionless coordinates, with the help of that the metric in new coordinate with $M_{0}=\frac{M}{r_{0}} =\frac{\tilde{l}^{2}-\epsilon}{2 \tilde{l}^{2}}$ and $\tilde{l}^{2}=\frac{l_{c}^{2}}{r_{0}^{2}}$ takes the following form,
 \begin{align}
 f(\tilde{r})&=1-\frac{2M_{0}}{\tilde{r}} +\epsilon \frac{\tilde{r}^{2}}{\tilde{l}^{2}}.
 \end{align}
 We will employ the usual numerical methodology to solve for the quasinormal modes of the system. In the usual quasinormal mode analysis, most of the cases the governing equation is Schr\"{o}dinger like wave equation. Therefore, by looking at the effective potential various analytical approximate methods have been developed over the years. However, for the present case, our equation is fourth order. Therefore, we will solve the problem numerically. In the transformed coordinate, the near horizon limit namely, in $\tilde{r}\rightarrow 1$, the master equation will take the following form,
 \begin{align}
 S^{''''}(\tilde{r}) +4 \frac{S'''(\tilde{r})}{(\tilde{r}-1)} + \left( \frac{9}{4} +\frac{W_{0}^{2}}{f'^{2}(1)} \right) \frac{S''(\tilde{r})}{(\tilde{r}-1)^{2}}-\left(\frac{2 m_{0}^{2}}{f'(1)} - \frac{f''(1)}{f'(1)}-\frac{3}{2} +\frac{l(l+1)}{f'(1)} -\frac{2W_{0}^{2}}{f'^{2}(1)} \right) \frac{S'(r)}{(r-1)^{2}} =0.
 \end{align} 
 Assuming the near horizon solution to be $S(\tilde{r})=(\tilde{r}-1)^{\nu}$ we get,
 \begin{align}
 \nu(\nu-1)\left\{\nu^{2}-\nu -2+\left( \frac{9}{4} +\frac{W_{0}^{2}}{f'^{2}(1)} \right)\right\}=0.
 \end{align}
 We have four roots which are $\nu=0,1$, and
 \begin{align}
 \nu^{\pm}&= \frac{1}{2}  \pm \frac{iW_{0}}{f'(1)}.
 \end{align}
 For the asymptotically flat Schwarzschild and Schwarzschild de Sitter black holes the above two imaginary roots will take the following forms respectively,
 \begin{align}
 &\nu^{\pm}_{Sch}= \frac{1}{2} \pm i W_{0}, \\
& \nu^{\pm}_{dS}= \frac{1}{2} \pm \frac{i W_{0}\tilde{l}^{2}}{\tilde{l}^{2}-1},
 \end{align}
 where, $``+$'' corresponds to outgoing and $``-"$ corresponds to ingoing mode near the horizon of the black hole. For very large values of $\tilde{l}^{2}$ we will have $\nu^{\pm}_{Sch} \sim \nu^{\pm}_{dS}$. From Eq. \ref{eq29}, we have only one free parameter in the theory which is $m_{0}^{2}$. In the following two sub-sections our goal would be to find out the behaviour of the Goldstone mode in the asymptotic region and the associated quasinormal frequencies.  
 
 \subsection{Quasinormal frequencies for asymptotically flat Schwarzschild black hole}
 
  This section will deal with numerically finding out the quasinormal frequencies for the asymptotically flat Schwarzschild black holes with a scalar hair as discussed in the previous sections.  We employed the method developed by Chandrasekhar and Detweiler \cite{Chandrasekhar1975} for finding the quasinormal frequencies. {The procedure is to integrate the equation numerically by using ``shooting" from both the end of the radial coordinate. Two asymptotic solutions, one near the horizon with the ingoing mode  $(\tilde{r}-1)^{\nu^{-}}$ and another one near the asymptotic infinity with the outgoing mode $(\tilde{r}-1)^{\nu^{+}}$ are matched at some intermediate point $\tilde{r}_{int}$. While doing this matching we need to make sure that at the matching point the two aforesaid solutions assume the same numerical value and their derivative or in other words, the Wronskian of the two solutions must vanish. Given the fixed boundary conditions near the horizon and at the infinity, the above matching condition will follow only for certain discrete but complex values of the frequency $W_0$. This gives us the required quasinormal frequencies. However, while doing this numerical analysis, we need to judiciously choose the value of the intermediate point and the numerical infinity $\tilde{r}_{inf}$ so that the frequency obtained remains stable. Within the range of $\tilde{r}_{inf} =(50,100)$ with a given value of the intermediate point $\tilde{r}_{int}\sim 7$ we found stable quasinormal frequencies. One of the main difficulties probably lies in the fact that our equation is higher derivative in nature, and we can not extract a well-defined potential which generically exits in the usual quasinormal mode analysis. Importance of the effective potential is that its nature along the radial coordinate provides valuable information while doing numerical quasi-normal mode analysis. Hence in addition to non-applicability of standard analytic WKB method which we have used already in the previous section, numerical analysis for present higher derivative equation also becomes non-trivial.} In fact, at both the boundaries, the general solution of our differential equation is a mixture of exponentially growing and exponentially decaying modes and one needs to choose the pure exponentially growing modes to calculate the quasinormal frequencies. Numerically, too large values of the radial coordinate attract contributions from unwanted exponentially suppressed modes and these may become significant after the integration. This gives rise to different values of frequency for different choices of infinity in the same mode. This problem is generally avoided by choosing small values of numerical infinities, but keeping large enough order in the series expansion. 
 \begin{table}[t]
 	\centering
 	\begin{tabular}{ |c|p{3cm}|p{3cm}|p{3cm}|  }
 		\hline
 		\hline
 		$$ & $l=1$ & $l=2$ & $l=3$ \\
 		\hline
 		$n$ &  ${W_0}_R-i{W_0}_I$ &  ${W_0}_R-i{W_0}_I$  &  ${W_0}_R-i{W_0}_I$  \\
 		\hline
 		$1$ & $0.728 - 0.218 i$ & $0.965- 0.242 i$ & $1.153- 0.206 i$ \\
 		$2$& $0.871 - 0.313 i$   & $1.444- 0.424 i$ & $1.721- 0.337 i$\\
 		$3$& $1.302 - 0.416 i$ & $1.628- 0.429 i$ & $2.132- 0.450 i$\\
 		$4$& $1.594 - 0.447 i$ & $2.038- 0.451 i$ & $2.436- 0.456 i$\\
 		$5$&$1.979- 0.456 i$ & $2.377- 0.465 i$ & $2.800- 0.470 i$\\
 		$6$&$2.336- 0.469 i$ & $2.742- 0.474 i$  & $3.151- 0.479 i$ \\
 		$7$& $2.704- 0.477 i$ & $3.102- 0.483 i$ & $3.510- 0.487 i$\\
 		$8$& $3.069- 0.486 i$ & $3.466- 0.490 i$ & $3.871- 0.494 i$\\
 		$9$&$3.437- 0.492 i$ & $3.831- 0.497 i$ & $4.233- 0.500 i$\\
 		$10$&$3.804- 0.498 i$ & $4.196- 0.503 i$  & $4.596- 0.506 i$ \\
 		\hline
 		\end{tabular}
 		\caption{Quasinormal frequencies of Schwarzschild black hole for overtone numbers $n=1$ to $10$ for $m_{0}^{2} = 10^{-6}$. Different multipole numbers ($l=1$ to $3$) corresponding to the different overtones are shown in the table.}
 		\end{table}
 		
 		\begin{table}[t]
 		\centering
 		\begin{tabular}{ |c|p{3cm}|p{3cm}|p{3cm}|  }
 		\hline
 		\hline
 		$$ & $l=1$ & $l=2$ & $l=3$ \\
 		\hline
 		$n$ &  ${W_0}_R-i{W_0}_I$ &  ${W_0}_R-i{W_0}_I$  &  ${W_0}_R-i{W_0}_I$  \\
 		\hline
 		$1$ & $0.770 - 0.208 i$ & $1.026- 0.219 i$ & $1.211- 0.194 i$ \\
 		$2$& $0.905 - 0.304 i$   & $1.466- 0.442 i$ & $1.756 - 0.332 i$\\
 		$3$& $1.331 - 0.418 i$ & $1.651- 0.410 i$ & $2.147 - 0.451 i$\\
 		$4$& $1.613 - 0.437 i$ & $2.055- 0.449 i$ & $2.451- 0.451 i$\\
 		$5$&$1.996- 0.453 i$ & $2.391- 0.461 i$ & $2.812- 0.468 i$\\
 		$6$&$2.350- 0.466 i$ & $2.754 - 0.472 i$  & $3.162- 0.477 i$ \\
 		$7$& $2.716- 0.475 i$ & $3.113- 0.481 i$ & $3.520- 0.486 i$\\
 		$8$& $3.081- 0.484 i$ & $3.476- 0.489 i$ & $3.879- 0.493 i$\\
 		$9$&$3.447- 0.491 i$ & $3.840- 0.496 i$ & $4.241- 0.499 i$\\
 		$10$&$3.813- 0.497 i$ & $4.205- 0.502 i$  & $4.604- 0.505 i$ \\
 		\hline
 	\end{tabular}
 	\caption{Quasinormal frequencies of Schwarzschild black hole for a different value of $m_{0}^{2} = 10^{-1}$.}
 \end{table}
 Keeping all these things in mind, in order to proceed, we first figure out the asymptotic behaviour of the solution for the Goldstone mode fluctuation $\pi$ in $r \rightarrow \infty$ limit. For asymptotically flat black holes, the master Eq.\ref{eq29} takes the following asymptotic form as
 \begin{align}
 S''''(\tilde{r})&+\frac{4}{\tilde{r}}S'''(\tilde{r}) + \left(W_{0}^{2}-2 m_{0}^{2} \right) S''(\tilde{r}) + \frac{2}{\tilde{r}} \left(W_{0}^{2}-2 m_{0}^{2} \right) S'(\tilde{r}) =0.
 \end{align}
 The corresponding general solution turns out to be
 \begin{align}
 S(\tilde{r}) \sim -\frac{B_{1}}{\tilde{r}} - B_{2} \frac{e^{-i\tilde{r}\sqrt{W_{0}^{2}-2 m_{0}^{2}}}}{\tilde{r}(W_{0}^{2}-2 m_{0}^{2} )} +B_{3} \frac{i e^{i\tilde{r}\sqrt{W_{0}^{2}-2 m_{0}^{2}}}}{2\tilde{r}(W_{0}^{2} - 2 m_{0}^{2})^{3/2}} +B_{4},
 \end{align}
 where, $B$'s are the integration constants fixed by the appropriate boundary conditions. The outgoing boundary condition is fixed by the mode corresponding to the coefficient of $B_3$, and the condition on the theory parameters should be $2 m_{0}^{2} = \frac{4\tilde{M}_{3}^{6}r_{0}^{2}}{c^{2}} < W_{0}^{2}$. Another important point we want to make at this point is that as the spacetime is asymptotically flat, the bare cosmological constant $\cal{J}$ satisfies
 \begin{align}
 \Lambda= \frac{3}{l_{c}^{2}} = \frac{1}{M_p^2}\left(\frac{\tilde{M}_{2}^{12}}{4 \tilde{M}_{1}^{8}} + \cal{J}\right) = 0,
 \end{align} 
 which immediately sets the bare value of the cosmological constant in term of our theoretical parameters. With this ingredient, we further proceed to solve for the quasinormal modes for a class of asymptotically flat Schwarzschild black holes. In the numerical integration method described above, we express the  solution for $S(\tilde{r})$ for finite but large $\tilde{r}$ as an infinite series as   
 \begin{align}
 S(\tilde{r})&= \frac{e^{i\tilde{r}\sqrt{W_{0}^{2}-2 m_{0}^{2}}}}{\tilde{r}} H(\tilde{r}) = e^{i\tilde{r}\sqrt{W_{0}^{2}-2 m_{0}^{2}}}\sum_{n=0}^{\infty} \frac{g_{n}}{\tilde{r}^{n+1}}. 
 \end{align}
 \begin{figure}[t] 
 	\begin{minipage}[c]{\textwidth}
 		\centering
 		\includegraphics[width=3.1in]{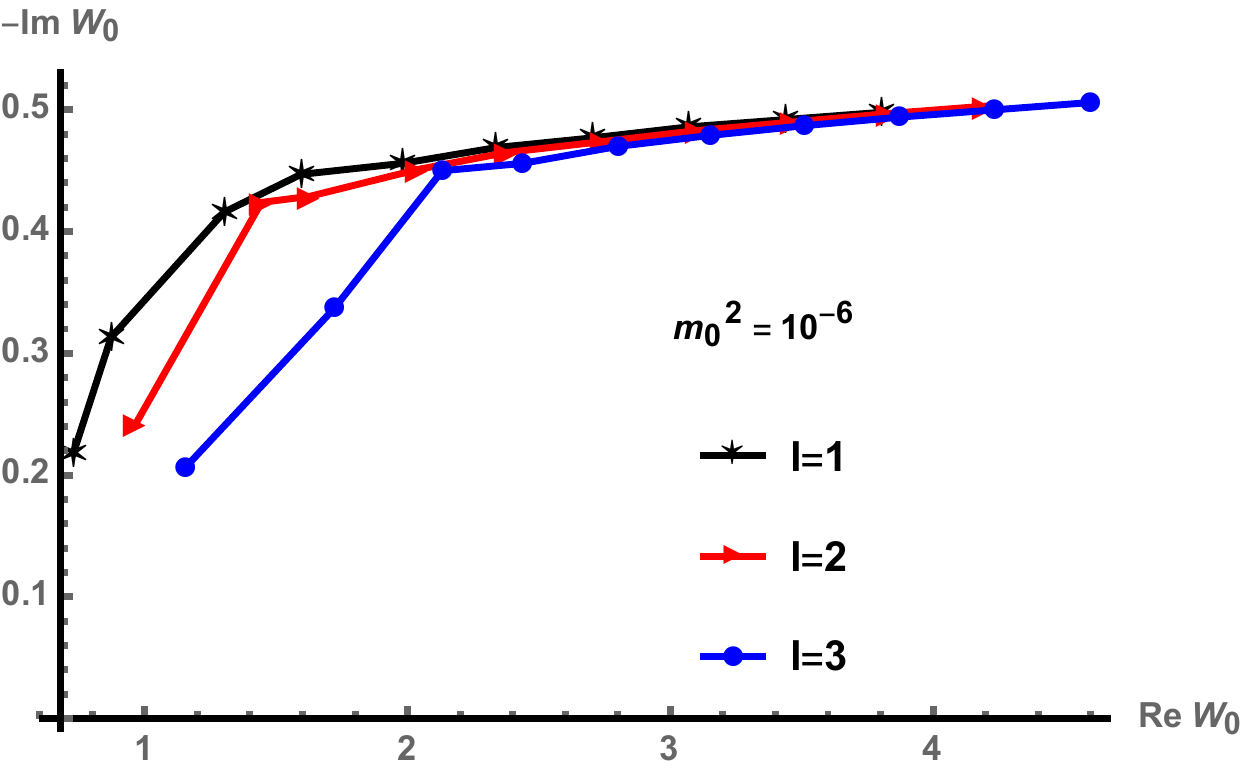}
 		\includegraphics[width=3.1in]{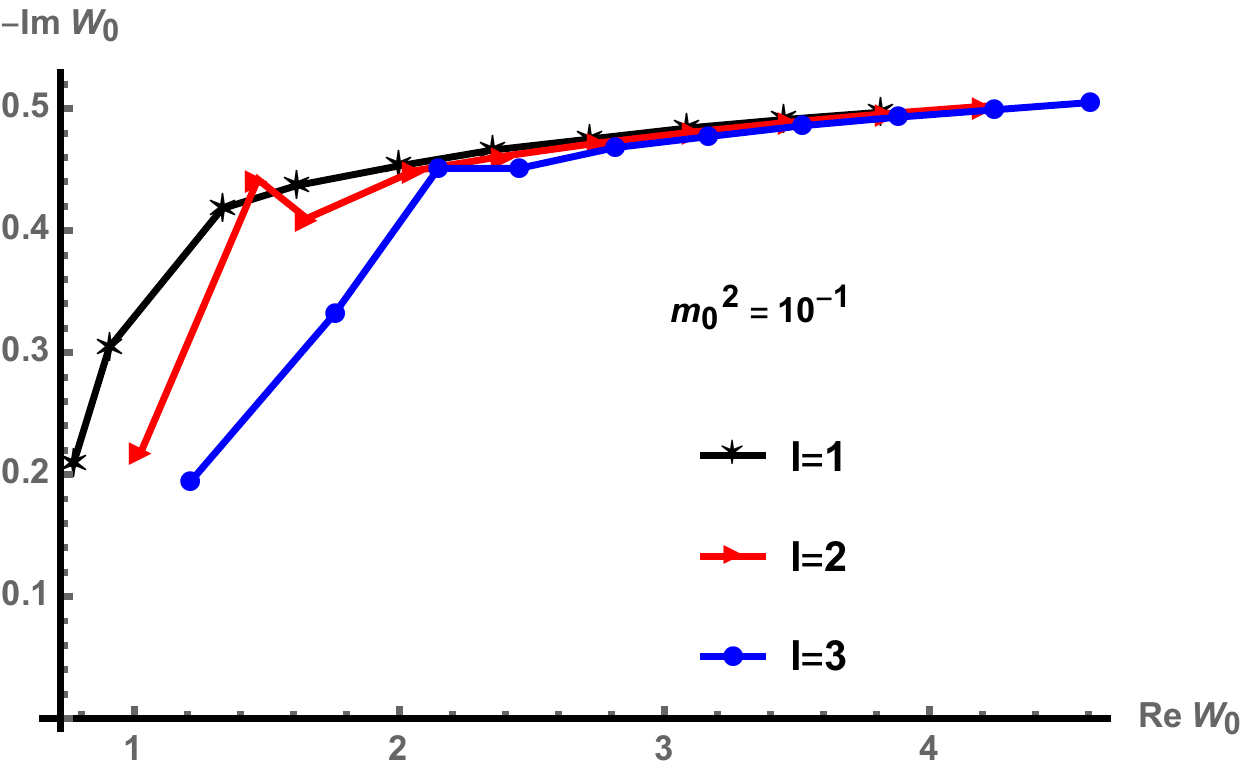}
 		\caption{Plot of the real and imaginary parts of $W_0$ for the Schwarzschild black hole for the first 10 overtones}
 	\end{minipage}
 \end{figure}
We have obtained the quasinormal frequencies using the numerical integration and by fixing different guess values of the frequency and found out stable values corresponding to different overtones. These are tabulated in Table (1) and (2). The nature of these frequencies thus obtained differs significantly from the nature of the pure Schwarzschild quasinormal frequencies. As the overtone number increases, the real oscillation frequency increases rapidly which is just the opposite behaviour when compared to the Schwarzschild black hole. On the other hand, it is known that the fundamental overtone will have the lowest damping and therefore the longest life. This feature is present here, however, the rate of increase of the imaginary part of the frequency in our case is slow compared to the Schwarzschild black hole. The real vs imaginary parts of the complex quasinormal frequencies for different $m_0^2$, overtones $n$ and multipole number $l$ are plotted in Fig. (3).

 \subsection{Quasinormal frequencies for Schwarzschild de Sitter black holes}
 
 We come to the final example of the applicability of our theory and will numerically find out the quasinormal frequencies for Schwarzschild de Sitter black holes. Similar to the discussion stated in detail for the asymptotically flat black holes, the asymptotic master Eq.\ref{eq29} in $r \rightarrow \infty$ for the present case takes the following form,
 \begin{align}
 S''''(\tilde{r})& +\frac{12}{\tilde{r}}S'''(\tilde{r})+ \left(34-\frac{2 m_{0}^{2} l^{2}_{c}}{\epsilon} \right)\frac{S''(\tilde{r})}{\tilde{r}^{2}} + \left(2-\frac{m_{0}^{2} l_{c}^{2}}{\epsilon}\right) \frac{8}{\tilde{r}^{3}} S'(\tilde{r})=0 .
 \end{align}
 Taking the asymptotic solution to be of the form, $S(\tilde{r}) = {\tilde r}^{{-}p}$, 
 the solution of the above equation for $\epsilon=-1$ (dS black hole), one finds $p = 0, 3$ and following conjugates 
 \begin{align*}
 p_{+}& =  \frac{1}{2} \left(3 + \sqrt{9-8 m_{0}^{2} \tilde{l}^{2}}  \right), \\
 p_{-} & = \frac{1}{2} \left(3 - \sqrt{9-8 m_{0}^{2} \tilde{l}^{2}}  \right).
 \end{align*}
 %
 The outgoing boundary condition sets an important constraint on our theoretical parameter, which is given by $9< 8 m_{0}^{2} \tilde{l}^{2}$. Similar to the Schwarzschild case discussed before we consider the following ansatz for the solution at the asymptotic infinity taking into account outgoing mode ($\tilde{r}^{-p_{-}} \equiv e^{-p_{-} \ln \tilde{r}}$), as
 \begin{align}
 S_{dS}(\tilde{r})&=  e^{-p_{-} \ln \tilde{r}} \sum_{n=0}^{\infty} \frac{g_{n}}{\tilde{r}^{n}} \equiv  e^{-p_{-} \ln \tilde{r}} H_{dS}(r)
 \end{align}
Applying the same numerical method we again solve for  $H_{dS}$ function. However, in order to solve numerically, we have taken into account the following constraints. The condition of outgoing mode at the horizon gives us lower bound on $m_{0}$ as
 \begin{align} \label{m0cond1}
 \frac{9}{8 \tilde{l}^{2}} & <m_{0}^{2} =\frac{6\tilde{M}_{3}^{6}}{X_{0}^{2}\tilde{l}^{2}}  M_{Pl}^{2} 
 \end{align} 
 {The above equation exhibits the dependence of our model parameter $m_{0}$ on the scalar hair parametrised by $X_{0}$. Now, since for our present study we are considering the Schwarzschild de Sitter black hole, assuming ${\cal J}=0$, the value of the de Sitter cosmological constant can be set by the non-zero scalar hair $X_0$. Therefore, combining the above condition Eq.\ref{m0cond1} with the value of the cosmological constant Eq.\ref{001a}, we can find the following constraint on our model parameter space as}, 
 \begin{align}
 \frac{3 M_{Pl}^{2}\tilde{M}_{2}^{12}}{32 \tilde{M}_{1}^{4}}< \tilde{M}_{3}^{6}
 \end{align}
 {Our goal is to establish a possible connection with the observations and to extract the possible allowed range of values of the model parameters $(\tilde{M}_{1}, \tilde{M}_{2}$ and $\tilde{M}_{3})$.
Therefore, we assume our present modified gravity model with the higher derivative scalar field Eq.\ref{action} as a phenomenological model for our Milky way black hole with the present value of the cosmological constant of our universe.    	
 Therefore, we identify our theoretical value of the cosmological constant $ \Lambda_{th} = {\tilde{M}_{2}^{12}}/({4 \tilde{M}_{1}^{8} M_p^2})$ to that of the observed value $\Lambda_{th}=\Lambda_{obs} \sim 10^{-47} GeV^{2} \sim 10^{-123}  M_{Pl}^{2}$, and also take the size of the supermassive black hole at the centre of the Milky way galaxy to be $r_{0} \approx 1.17 \times 10^{10} m$. With those values one can arrive at the following constraint on our theory parameters,} 
 \begin{align}
 1.6 \times 10^{-53} r_{0}^{2} m^{-2} < & m_{0}^{2}  \Rightarrow 9.45 \times 10^{-85} GeV^{2} < \frac{\tilde{M}_{3}^{6}}{\tilde{M}_{1}^{4}}   \\
  \Rightarrow 0.1875 c^{2} < & \tilde{M}_{3}^{6}  M_{Pl}^{2}  \\
 \Rightarrow 0.375  < & \frac{\tilde{M}_{3}^{6}}{\tilde{M}_{1}^{4} \Lambda} .
 \end{align}
For the above constraint equation, we consider $m_{0}=10^{-6}$.
 \begin{table}[t]
 	\centering
 	\begin{tabular}{ |c|p{3cm}|p{3cm}|p{3cm}|  }
 		\hline
 		\hline
 		$$ & $l=1$ & $l=2$ & $l=3$ \\
 		\hline
 		$n$ &  ${W_{0}}_R-i{W_{0}}_I$ &  ${W_{0}}_R-i{W_{0}}_I$  &  ${W_{0}}_R-i{W_{0}}_I$  \\
 		\hline
 		$1$ & $0.743 - 0.261 i$ & $0.950 - 0.217 i$ & $1.151 - 0.203 i$ \\
 		$2$&    & $1.453 - 0.420 i$ & $1.653 - 0.334 i$\\
 		\hline
 		\end{tabular}
 		\caption{Quasinormal frequencies of Schwarzschild-dS black hole  for $m_{0}^{2} = 10^{-6}$ and $\tilde{l}^{2}=10^{8}$}
 		\label{tab3}
 		\end{table}
 		\begin{table}[t]
 		\begin{tabular}{ |c|p{3cm}|p{3cm}|p{3cm}|  }
 		\hline
 		\hline
 		$$ & $l=1$ & $l=2$ & $l=3$ \\
 		\hline
 		$n$ &  ${W_{0}}_R-i{W_{0}}_I$ &  ${W_{0}}_R-i{W_{0}}_I$  &  ${W_{0}}_R-i{W_{0}}_I$  \\
 		\hline
 		$1$ & $0.818 - 0.243 i$ & $1.014 - 0.204 i$ & $1.208 - 0.192 i$ \\
 		$2$&    & $1.495 - 0.417 i$ & $1.690 - 0.333 i$\\
 		\hline
 	\end{tabular}
 	\caption{Quasinormal frequencies of Schwarzschild-dS black hole  for $m_{0}^{2} = 0.1$ and $\tilde{l}^{2}=10^{8}$}
 	\label{tab4}
 \end{table}
As an example, choosing following set of numerical values of the parameters $\tilde{M}_{1} \sim 10^{-7/4}$ GeV and $\tilde{M}_{2} \sim 10^{-5}$ GeV we get the condition 
 \begin{align}
 m_{0}^{2} & =4 \frac{\tilde{M}_{3}^{6} \tilde{M}_{1}^{4} \Lambda}{\tilde{M}_{2}^{12}}  M_{Pl}^{2} r_{0}^{2} \sim 4 \times 10^{6} ( M_{Pl}^{2} \tilde{M}_{3}^{6} GeV^{-6} r_{0}^{2})
 \end{align}
which leads to $m_{0}^{2} \sim 10^{-6}$ in unit of $r_0$ choosing  $\tilde{M}_{3}=10^{-2} GeV$. Our numerical results for the quasinormal modes are given in the table \ref{tab3},\ref{tab4} for two set of $m_0^2$. For this case, our numerical analysis was not stable enough to give the quasinormal modes for higher overtones. However, qualitative behaviour of the modes remains same as that of the Schwarzschild black hole case discussed earlier.

 %

	\section{Conclusion}
		Effective field theory is a powerful approach towards understanding the low energy behaviour for a wide range of physical phenomena. In the present paper, we applied this approach towards understanding black holes in a model-independent manner. In order to probe any physical system, the behaviour of fluctuation in a certain background under consideration is most important. In this paper, we consider black hole space time with hair as a particular type of background whose properties can only be understood by looking at the nature of fluctuation around it. The conventional effective field theory approach deals with writing down the theory of background itself in terms of fundamental fields. In the present context, the approach was to consider prior existence of a background of interest and then write down the most general theory for the fluctuations in the given background based on the symmetry. This is the approach which has been successfully applied in the inflationary cosmology which is popularly known as an effective theory of inflation. In this paper, we apply the same technique in the background of spherically symmetric black holes with hair which also enjoys the same symmetry. As mentioned earlier our current investigation is motivated by the following question: {\em does the effective theory of black holes provide any information about the possible existence of hair?} 
		
We have first written down the most general model-independent effective Lagrangian for the fluctuation in a given hairy black hole background. For our detailed analysis, we have considered an asymptotically flat and de-Sitter black hole background. In which the background cosmological constant is assumed to be generated from the hair. Generally, the behaviour of fluctuation encodes important information about background hair.
Therefore, in order to understand the behaviour of the fluctuation we have chosen a particular set of effective theory parameters. By using the sixth order WKB approximation, associated with those fluctuations we have computed the quasinormal modes which appeared to carry different features when compared with that of usual black holes quasinormal modes.
In general, for the four-dimensional  Schwarzschild black hole in the asymptotically flat/dS background, the real oscillation frequency of the quasinormal modes decreases and imaginary part of the frequency increases with the increase of the overtone number (n) while the multipole number (l) is kept fixed. Interestingly what came out from our quasinormal mode analysis for the effective field theory fluctuation is that both the real and imaginary frequencies increase with increasing overtone number.  
Motivated by our effective field theory analysis, we also constructed a class of higher derivative scalar field theory. For this theory also we confirmed the aforementioned interesting behaviour of the quasinormal modes.

\section{Acknowledgements}
Sumeet Chougule is thankful to Krishnamohan Parattu, Pankaj Saha and Wadbor Wahlang for many useful discussions and he is also thankful for the hospitality provided by the IIT Guwahati. Sumeet Chougule is funded by Conicyt grant 21181211. The Centro de Estudios Científicos (CECs) is funded by the Chilean Government through the Centers of Excellence Base Financing Program of Conicyt.

	
%
%

\end{document}